\documentclass[fleqn,usenatbib]{mnras}

\usepackage[usenames,dvipsnames]{xcolor}

\hypersetup{
    colorlinks = true,
    citecolor = {MidnightBlue},
    linkcolor = {BrickRed},
    urlcolor = {BrickRed}
}

\usepackage{mathptmx}
\usepackage{amsmath}
\usepackage{amssymb}

\usepackage[english]{babel}
\usepackage{txfonts}

\usepackage{graphicx}

\usepackage{booktabs}
\usepackage{multirow}
\usepackage{float}
\usepackage{placeins}
\usepackage{balance}
\usepackage{cuted}

\usepackage{geometry}
\usepackage{tkz-graph}
\usetikzlibrary{arrows}

\newcommand{\be}{\begin{equation}}
\newcommand{\ee}{\end{equation}}
\newcommand{\bea}{\begin{eqnarray}}
\newcommand{\eea}{\end{eqnarray}}

\newcommand{\mhalo}{M_{\rm h}}
\newcommand{\mstar}{M_\star}
\newcommand{\sfr}{\psi_{\rm SFR}}
\newcommand{\corr}{}

\title[Galaxy-Halo Model for Multiple Tracers]{A Galaxy-Halo Model for Multiple Cosmological Tracers}

\author[P. Bull]{
Philip Bull$^{1, 2}$\thanks{E-mail: philip.j.bull@jpl.nasa.gov}
\\
$^{1}$California Institute of Technology, Pasadena, CA 91125, USA \\
$^{2}$Jet Propulsion Laboratory, California Institute of Technology, 4800 Oak Grove Drive, Pasadena, California, USA
      }

\date{Accepted 28 April 2017. Received 27 April 2017; in original form 1 November 2016}
\pubyear{2017}

\begin{document}
\label{firstpage}
\pagerange{\pageref{firstpage}--}
\maketitle

\begin{abstract}
The information extracted from large galaxy surveys with the likes of DES, DESI, Euclid, LSST, SKA, and WFIRST will be greatly enhanced if the resultant galaxy catalogues can be cross-correlated with one another. Predicting the nature of the information gain, and developing the tools to realise it, depends on establishing a consistent model of how the galaxies detected by each survey trace the same underlying matter distribution. Existing analytic methods, such as halo occupation distribution (HOD) modelling, are not well-suited for this task, and can suffer from ambiguities and tuning issues when applied to multiple tracers.
{\corr In this paper, we take the first steps towards constructing an alternative} that provides a common model for the connection between galaxies and dark matter halos across a wide range of wavelengths (and thus tracer populations).
This is based on a chain of parametrised statistical distributions that model the connection between (a) halo mass and bulk physical properties of galaxies, such as star-formation rate; and (b) those same physical properties and a variety of emission processes.
The result is a flexible parametric model that allows analytic halo model calculations {\corr of 1-point functions} to be carried out for multiple tracers, as well as providing semi-realistic galaxy properties for fast mock catalogue generation.
\end{abstract}


\section{Introduction}


The viability of large-scale structure surveys as a cosmological probe rests on our ability to understand the connection between galaxies and the dark matter distribution that they inhabit. Galaxies are the luminous `tracers' of the dark matter field that we actually observe, while the clustering of the invisible dark matter, much of which has collapsed into discrete halos, bears the bulk of the cosmological information. Without a sufficiently accurate model of their relationship, forthcoming galaxy surveys are limited in their ability to return accurate, precision cosmological constraints \citep{2014MNRAS.444..476R, 2015MNRAS.448.1389C}. As a result, there has been a concerted effort to develop theoretical models of the galaxy-dark matter connection, and validate them against simulations and existing observations across a wide range of frequencies \citep[e.g.][]{2003ApJ...593....1B, 2005ApJ...630....1Z, 2009ApJ...707..554Z, 2009MNRAS.392.1080S, 2011ApJ...728..126W}. 

A defining feature of forthcoming surveys is their coverage of large fractions of the sky. A number of the datasets will therefore cover substantially overlapping cosmological volumes, enabling cross-correlation analyses to be performed. The relationship between different galaxy samples (often seen at different wavelengths) yields valuable additional information beyond what a single tracer can provide, both about the galaxies themselves and the large-scale matter distribution \citep[e.g.][]{2012MNRAS.422.2904G}. Cross-correlations can be used to side-step the cosmic variance limit on certain observables, for example \citep{McDonald:2008sh}; to defeat otherwise difficult systematic effects \citep{Camera:2016owj}; or to uncover the physical properties and formation processes of the galaxies themselves \citep{2007arXiv0709.1159J}.

While extremely promising, such analyses also require the application of a suitable galaxy-halo model -- with the added complication that it must be {\it consistent} between tracers. This is challenging for methods designed primarily for the single-tracer case. Existing multi-wavelength data are often limited in depth, or by sample variance, making it hard to empirically calibrate the models against two or more tracers simultaneously, while including the necessary correlation information between tracers.

In this paper, we construct a modular analytic model for the connection between galaxies and halos that provides consistent predictions across multiple wavelengths and tracer populations. We achieve this by using empirical scaling relations to connect the host halo mass to bulk physical properties of the galaxies, such as stellar mass and star-formation rate. These relationships are described in a statistical manner, by constructing appropriately-parametrised probability distributions around the mean scaling relations (which are also parametrised). The physical properties are then related to a variety of emission processes by another set of scaling relations, each of which is relevant to some wavelength regime. Correlations between different types of emission (and thus different frequency bands) arise because of their shared dependence on the basic physical properties of the galaxy.

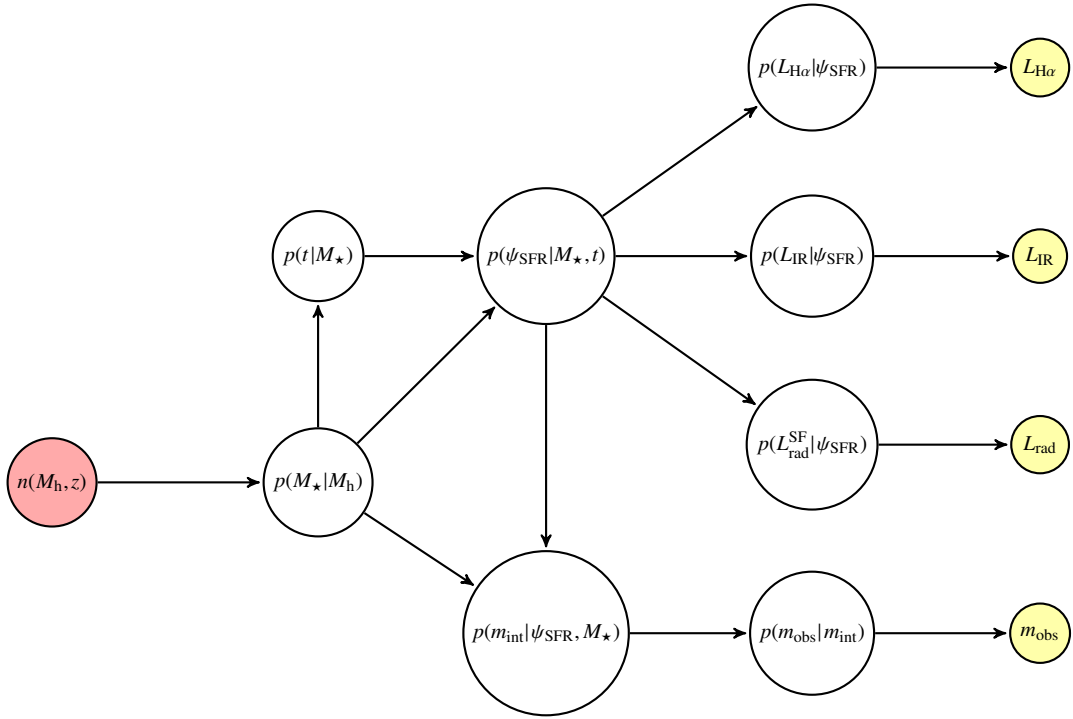
\begin{figure*} \centering
\begin{tikzpicture}[->,>=stealth',shorten >=1pt,thick,every node/.style={draw=black,circle,align=center}]
\SetGraphUnit{4} 
\GraphInit[vstyle=Normal] 
\SetVertexNormal[Shape=rect,MinSize=1cm,LineWidth=1pt]
\tikzset{VertexStyle/.append every node/.style={draw=black,rect}} 

  \node[fill={rgb:red,1;white,2}] (mhalo) at (-1,0) {$n(\mhalo, z)$};
  \node (mstar) at (2.5,0) {$p(\mstar|\mhalo)$};
  
  \node (sftype) at (2.5,3) {$p(t\,|\mstar)$};
  \node (sfr) at (5.5,3) {$p(\psi_{\rm SFR}|\mstar, t)$};
 
 \node (halpha) at (9,5.5) {$p(L_{{\rm H}\alpha}|\sfr)$};
 \node (ir) at (9,3) {$p(L_{\rm IR}|\sfr)$};
 \node (sfradio) at (9,0.5) {$p(L^{\rm SF}_{\rm rad}|\,\sfr)$};
 \node (optical) at (5.5,-2) {$p(m_{\rm int}|\,\sfr, \mstar)$};
 \node (optobs) at (9,-2) {$p(m_{\rm obs}|\,m_{\rm int})$};
 
 \node[fill={rgb:yellow,1;white,2}]
    (Lhalpha) at (12,5.5) {$L_{{\rm H}\alpha}$};
 \node[fill={rgb:yellow,1;white,2}]
    (Lir) at (12,3) {$L_{\rm IR}$};
 \node[fill={rgb:yellow,1;white,2}]
    (Lradio) at (12,0.5) {$L_{\rm rad}$};
 \node[fill={rgb:yellow,1;white,2}]
    (Lopt) at (12,-2) {$m_{\rm obs}$};
 
  \draw[->] (mhalo) to (mstar);
  \draw[->] (mstar) to (sfr);
  \draw[->] (mstar) to (sftype);
  \draw[->] (sftype) to (sfr);
  \draw[->] (mstar) to (optical);
  
  \draw[->] (sfr) to (halpha);
  \draw[->] (sfr) to (sfradio);
  \draw[->] (sfr) to (ir);
  \draw[->] (sfr) to (optical);
  
  \draw[->] (halpha) to (Lhalpha);
  \draw[->] (ir) to (Lir);
  \draw[->] (sfradio) to (Lradio);
  \draw[->] (optical) to (optobs);
  \draw[->] (optobs) to (Lopt);
\end{tikzpicture}

\caption{Graph of the probabilistic model linking input parameters (halo mass $\mhalo$ and redshift $z$, drawn from a halo mass function) to observable luminosities/magnitudes in various bands. Only the radio and optical bands are considered in this paper; the IR and H$\alpha$ bands are shown for illustration.}
\label{fig:model}
\end{figure*}

The `building blocks' of the model -- a chain of interconnected parametric statistical distributions (Fig.~\ref{fig:model}) -- are mostly based on established relations that are found in the literature, such as the star-forming main sequence, or stellar mass-halo mass relation. By {\it jointly} fitting the parameters of the model to various datasets, rather than relying on empirical calibrations of individual components from the literature, we can ensure that the components of the model all connect together in a consistent way. Each component can be replaced or upgraded as more realistic models become available, and uncertainties on model parameters can be straightforwardly propagated through the full system to determine their effects on observables, using standard Monte Carlo techniques or even analytic marginalisation.

The model is differentiated from similar approaches in the literature by its analytic construction, meaning that it can be used for rapid parameter estimation and forecasting, as well as the more usual task of painting galaxy properties onto dark matter-only simulations to create mock catalogues. A reference implementation, written in Python, is made publicly-available with this paper.

As a proof of concept, in this paper we will construct a model for optical and radio continuum emission from `normal' (non-AGN) galaxies at $z \simeq 0$, and calibrate it off multi-wavelength luminosity function data ({\corr plus information from a} semi-analytic simulation for one model component). Simple extensions of the model can be made to generalise it to higher redshifts and several other tracer populations. Other novel features include a new set of analytic fitting functions that connect optical magnitudes to star-formation rate and stellar mass, and a simple dust attenuation model.


{\corr This represents only an initial step in the development of a fully viable multi-tracer galaxy-halo model. Our ultimate goal is to construct a halo model with sufficient complexity to describe and predict the joint 1-point (luminosity function) and 2-point (clustering) statistics of multiple galaxy populations observed across the full range of wavebands covered by forthcoming large cosmological surveys like Euclid, LSST, SKA, and WFIRST. As discussed below, this will require significantly more complexity than the single-tracer HOD models currently in use, but hopefully significantly less complexity than semi-analytic modelling or hydrodynamical simulations. This is a reasonable goal, as the aim is only to accurately describe the general properties of {\it populations} of galaxies, rather than detailed properties of individual galaxies.}

The paper is organised as follows. In Sect.~\ref{sec:popmodels} we briefly review various approaches that have been used to model galaxy populations. We then define the components of our model in Sect.~\ref{sec:model}, and calibrate it against simulations and observations in Sect.~\ref{sec:params}. We also analyse how well the model fits the input data, and discuss some of the other observables it can predict. We then conclude in Sect.~\ref{sec:discussion} with a discussion of the model's limitations, and ways that it could be improved and extended.

We denote the natural logarithm by $\log$ and the base-10 logarithm by $\log_{10}$ throughout, and assume the \citet{2016A&A...594A..13P} best-fit $\Lambda$CDM cosmology with $h=0.67$ and $\Omega_{\rm M}=0.32$.

\section{Galaxy population modelling} \label{sec:popmodels}

In this section, we briefly review some of the methods for modelling galaxy populations. These generally fall into one of three main categories:
\begin{itemize}
 \item Hydrodynamical simulations;
 \item Semi-analytic models;
 \item Halo model approaches.
\end{itemize}

{\it Hydrodynamical simulations} \citep[e.g.][]{2014MNRAS.445..581H, 2014Natur.509..177V} are typically used to make samples of galaxies with highly realistic properties. By explicitly modelling the physical processes relevant to galaxy formation and evolution, very detailed simulations can be constructed, albeit at very high computational expense. This typically limits the simulations to small spatial volumes, so that sufficiently high spatial resolutions can be achieved to model the necessary processes. `Sub-grid' models can be used to incorporate processes that happen on unresolvable length scales, although these typically introduce a number of free parameters that must be tuned to match observations. While useful for building a detailed understanding of galaxies, hydrodynamical simulations are generally too expensive to use for applications such as creating large numbers of mock galaxy catalogues for the analysis of large-scale structure surveys.

{\it Semi-analytic models} (SAMs) are commonly used to populate dark matter-only N-body simulations with galaxies. They also work by explicitly modelling the various physical processes responsible for determining observable galaxy properties, subject to some simplifying assumptions designed to reduce computational overhead \citep{1999MNRAS.310.1087S, 2000MNRAS.311..576K, 2005Natur.435..629S, 2012NewA...17..175B, 2013PASA...30...30B}. The simplified physical models in SAMs often have tunable parameters that can be calibrated against observations. While intended to be cheaper than hydrodynamical simulations but much more realistic than halo model approaches, SAMs are still computationally intensive due to the need to repeatedly solve systems of differential equations to determine the properties of each galaxy. This makes them unwieldy for performing Monte Carlo parameter studies for example, which require many different realisations of the galaxy population to be constructed for different sets of parameter values \citep[although such studies have been attempted; see e.g.][]{2009MNRAS.396..535H}. Minimal SAMs do exist, which can be used to more rapidly model some subset of galaxy properties \citep[e.g.][]{2016arXiv160903956C}.

{\it Halo model approaches} are a much lighter alternative than either of the previous two methods. These work by assigning galaxy properties to dark matter halos, using a set of scaling relations that depend on properties such as the halo mass. Halo models have the advantage of often being simple enough to evaluate analytically, making them suitable for cheaply populating mock catalogues \citep[e.g.][]{2013MNRAS.428.1036M, 2016MNRAS.459.2118K}, or performing statistical analyses \citep[e.g.][]{2014MNRAS.444..476R}.

Halo model methods come in a number of different flavours. Halo occupation distribution \citep[HOD;][]{2000MNRAS.318.1144P, 2002ApJ...575..587B, 2003ApJ...593....1B} modelling predicts the mean number of galaxies of a given type that reside in a halo of given mass, $\langle N \rangle(\mhalo)$. HODs are typically calibrated empirically, with the resulting fits being extrapolated to model the distribution of the same type of galaxies in future surveys. While HODs can be simultaneously defined for multiple galaxy populations \citep[e.g.][]{2010ApJ...713..558K, 2011ApJ...726...83M, 2013MNRAS.428.2548K}, the correlation between each pair of populations must be modelled (and calibrated) separately, e.g. using the observed cross-correlation function. Ambiguities may also arise when trying to ascertain which galaxies appear in more than one survey -- HODs calibrated at different wavelengths can predict different numbers of galaxies to exist in a given halo, so some galaxies will be `missing' in one band, but not in another.

An alternative approach is to construct conditional luminosity functions \citep[CLFs;][]{Yang:2002ww, 2004MNRAS.353..189V, Cooray:2005mm}, which describe the expected luminosity distribution of galaxies hosted in halos of a given mass. These methods are again calibrated empirically against a given population of galaxies, but can be extended to other wavelengths by assuming an appropriate spectral energy distribution (SED). Correlations between different types of tracer must again be added by hand though.

A unified framework for implementing halo model methods such as these is presented in \citet{2016arXiv160604106H}.

Several recent works have explicitly sought to model galaxy populations across multiple wavelengths. Some are empirical, in the sense that they create realisations of mock galaxy samples that are consistent with a set of observed luminosity functions (LFs) by construction \citep{2009A&A...504..359J, 2016arXiv160605354S}. The data for these are generally obtained from deep but narrow surveys (e.g. COSMOS), so the redshift dependence and faint end of the luminosity function are well measured, but large-scale clustering information tends to be minimal. A related approach is to use machine learning techniques to create a predictive model of a multi-wavelength galaxy sample, again by training it off an existing galaxy catalogue \citep{2013ApJ...772..147X}.

Alternatively, one can take scaling relations between galaxy properties (e.g. star-formation rate and stellar mass) and combine them to reproduce observables. This is the basic approach taken in this paper, and {\corr was also used by \citet{vandenBosch:2002zn, 2009MNRAS.392.1080S, 2016RAA....16h..13L, 2016arXiv160605354S}, and others}. This has the advantage of allowing several different observables to be constructed from a common set of components, and is justified by the empirical discovery of a number of appropriate scaling relations. A possible problem is inconsistency between the various components, which may be calibrated off different datasets with different assumptions. We will avoid this problem here by performing global model fits, rather than calibrating components individually.


\section{Model definition} \label{sec:model}

In this section, we specify each component of the model in detail, according to the basic structure illustrated in Fig.~\ref{fig:model}. {\corr The rationale behind this structure is that any joint probability distribution (e.g. of galaxy properties) can be decomposed into a chain of conditional distributions of the form $p(x_1, x_2,\ldots x_n) = p(x_1) \cdot p(x_2 | x_1) \ldots \cdot p(x_n | x_1, x_2,\ldots x_{n-1})$. Subsets of the conditional distributions can be grouped together by rewriting them as joint distributions over a subset of the variables, which in many cases can be approximated using simple analytic distributions, or even products of univariate distributions if the variables are essentially independent. This can vastly simplify the problem of modelling the full multivariate distribution if an appropriate simplifying restructuring can be found. The use of conditional distributions as tractable `building blocks' of complicated joint distributions is widespread in other contexts, e.g. in Gibbs sampling methods \citep{casella1992}.}

{\corr Using this approach, we apply known empirical relations between galaxy properties, plus some simple physical reasoning, to arrive at the simplified probabilistic model shown in Fig.~\ref{fig:model}. This is by no means unique, and in its present form is unable to account for some known correlations between galaxy properties (e.g. galaxy colours). These limitations would likely be exacerbated when considered clustering information, and so the model structure will need to be revisited in future. Nevertheless, as we will show below, the current structure is sophisticated enough to be able to model the joint luminosity functions of multiple galaxy populations at radio and optical wavelengths -- a non-trivial application that would normally be the preserve of SAMs.}

\subsection{Halo mass function}

The initial input to the model is a distribution of dark matter (DM) halos in mass, position, and redshift. This is represented by the halo mass function,
\be
n(\mhalo, z) \equiv \frac{dN(z)}{d\log \mhalo\, dV\, dz},
\ee
where $\mhalo$ is the halo mass and $V$ is the comoving volume. This distribution ultimately sets the overall abundance and clustering properties of galaxies, and the dependence on the underlying cosmological model. As a simplifying assumption, we will assume that DM halos are characterised solely by their mass and redshift, ignoring other intrinsic properties such as angular momentum and concentration. Furthermore, we will make no explicit distinction between parent and sub-halos, although in reality some galaxy properties do depend on this. Finally, we will assume that each halo (or sub-halo) hosts precisely one galaxy -- there are no empty halos, and no multiply-occupied ones (in contrast to HOD, where satellites are explicitly modelled).

In most of what follows we will use an analytic form for the halo mass function, but it is important to note that the model can be applied to {\it any} representation of a set of DM halos. For example, the chain of statistical distributions shown in Fig.~\ref{fig:model} can just as well be applied to a DM halo catalogue taken from an N-body simulation, and used to populate mock catalogues of galaxies using a Monte Carlo method. 

For the analytic calculations in the remainder of the paper, we will adopt the \cite{2008ApJ...688..709T} mass function,
\bea
n(\mhalo, z) &=& -f(\sigma) \frac{\rho_{\rm m}}{\mhalo} \frac{d\log \sigma(\mhalo)}{d \log \mhalo} \\
f(\sigma) &=& A \left [ 1 + \left ( \frac{\sigma(\mhalo)}{b} \right )^{-a}\right ] \exp \left ({-{c/\sigma^2}} \right ),
\eea
where $\rho_m$ is the background matter density and $\sigma(M)$ is the rms density fluctuation within a sphere of radius $R = (3 M / 4\pi \rho_m)^\frac{1}{3}$, from linear theory. The dependence on cosmological parameters is implicit in $\sigma(M)$, an integral over the matter power spectrum, $P(k)$, which we take from CAMB \citep{2000ApJ...538..473L}. The best-fit parameters, calibrated against simulations, are $A = 0.186$, $a = 1.47$, $b = 2.57$, and $c = 1.19$, all at $z=0$.

For calculations of the clustering, one must also specify a halo bias function, $b(\mhalo)$. The (Eulerian) halo bias is
\be
b(\mhalo) = 1 + (a \nu_1 - 1) / \delta_1 + \frac{2 p / \delta_1}{1 + (a\nu_1)^p},
\ee
where $\nu_1 = [\delta_1 / \sigma(\mhalo)]^2$ and $\delta_1 \approx \delta_c$.

\subsection{Stellar mass -- halo mass relation}

As a first step, it is necessary to link dark matter halos to their baryonic content. We model this through the relationship between halo mass and stellar mass, as described by the conditional mass function (CMF). This is related to the conditional probability to find a stellar mass of $\mstar$ in a halo of mass $\mhalo$: $p(\mstar | \mhalo)$.

Halo mass and stellar mass are expected to be strongly related as a simple consequence of hierarchical structure formation; as more massive halos collapse, they trap a correspondingly larger mass of baryonic matter, some fraction of which forms stars as the gas settles and cools in the centre of the halo potential \citep{1978MNRAS.183..341W}. A relatively tight relationship between the two is indeed seen observationally \citep{2007ApJ...671..153Y, 2010ApJ...710..903M, 2010ApJ...717..379B}, and is well fit by a broken powerlaw,
\be
\overline{\mstar}(\mhalo) = \frac{2 \mhalo\, A_\star}{\left(\frac{\mhalo}{M^\star_1}\right)^{-\beta_\star} + \left(\frac{\mhalo}{M^\star_1}\right)^{\gamma_\star}}, \label{eq:msmh}
\ee
where $A_\star$ is the overall normalisation, $M^\star_1$ is a mass scale, and $\beta_\star$ and $\gamma\star$ are the low- and high-mass end slopes respectively.
The conditional pdf is modelled as log-normal,
\be
p(\mstar | \mhalo) = \frac{1}{\sqrt{2\pi} \mstar \sigma_{\mstar}}\exp\left( -\frac{\log^2 (\mstar / \overline{\mstar})}{2 \sigma_{\mstar}^2}\right ),
\ee
where $\sigma_{\mstar}$ is the scatter, modelled by \cite{2010ApJ...710..903M} as
\be
\frac{\sigma_{\mstar}}{\log 10} = \sigma^\star_\infty + \sigma^\star_1 \left (1 - \frac{2}{\pi} \arctan \left [\xi_\star\, \log_{10}(\mhalo/M^\star_2) \right] \right ). \label{eq:mstar_sigma}
\ee
While \citet{2010ApJ...710..903M} provide separate fits for both central and satellite galaxies, we use only the central relation as, for simplicity, our model ignores this distinction.

\subsection{Star-formation rate -- stellar mass relation} \label{sec:sfrmstar}

The star-formation rate (SFR), $\sfr$, is a key quantity that predicts many of a galaxy's other observable properties. It is thought to depend on a number of factors, including the availability of cold molecular gas (the raw material of star formation); the recent merger history of the galaxy (mergers disturb cold gas, inducing its collapse to form stars); and quenching processes such as feedback from accretion onto the central supermassive black hole or supernovae. See \cite{2015MNRAS.453.4337S} and references therein for a more thorough overview.

As we are only interested in the aggregate properties of a large galaxy sample, we adopt stellar mass as a sufficiently robust predictor of the SFR. A connection between the two is well established -- a {\it star-forming main sequence} (SFMS) is found in the $\sfr-\mstar$ plane, along which the bulk of actively star-forming galaxies lie \citep{2007ApJ...670..156D, 2007A&A...468...33E, 2007ApJ...660L..43N, 2007ApJS..173..267S, 2010ApJ...721..193P, 2011ApJ...730...61K, 2013MNRAS.431..648W}. A second population of {\it quiescent} or {\it passive} galaxies lies below the SFMS, which typically contain older, redder stars, and less (but not necessarily negligible) star formation. This population may be arranged in a similar, but more scattered, sequence to the SFMS, or else in a much looser, relatively uniform distribution in SFR below the main sequence; see \cite{2011MNRAS.416.1566L} for a comparison of semi-analytic model predictions. Other populations may be identified in the $\sfr-\mstar$ plane, such a starburst galaxies (which lie above the SFMS), but we will work only with the simple active vs. passive categorisation here.

Motivated by the above, we will model the distributions of passive and SFMS galaxies separately. The first step is to assign a probability that a galaxy will belong to one population or the other. Several studies have attempted to estimate the fraction of passive galaxies from recent surveys \citep[e.g.][]{2009A&A...501...15F, 2010ApJ...721..193P, 2013ApJ...767...50M}; we adopt a modified version of the relation from \cite{2013ApJ...770...57B}, which they attributed to \cite{2011ApJ...739...24B}:
\be \label{eq:fpassive}
f_{\rm passive}(\mstar) = c + \frac{1 - c}{1 + \left ( \frac{\mstar}{10^{\,\alpha_f} M_\odot} \right )^{\,\beta_f}},
\ee
where we reparametrise $c \equiv \frac{1}{2} (1 + \tanh \zeta_f)$ to ensure that $f_{\rm passive} \in [0, 1]$. Factors other than stellar mass can also be important for determining whether a galaxy is passive or not \citep[e.g. environment; see][]{2010ApJ...721..193P}. We assume that these factors have been marginalised over here, leaving $\mstar$ as the only relevant variable. The resulting pdf, $p(t | \mstar)$, where $t$ denotes the galaxy type, is a uniform distribution partitioned at $f_{\rm passive}$.

For the SFMS, we adopt a mean relation with a similar form to the one found in \cite{2013MNRAS.431..648W},
\be \label{eq:meansfms}
\overline{\psi}_{\rm SFR}(\mstar) = 10^{\alpha_{\rm SFMS}} \left(\frac{\mstar}{10^{10} M_\odot}\right)^{\beta_{\rm SFMS}}\, M_\odot {\rm yr}^{-1}.
\ee
The scatter in the SFMS is incorporated by modelling it as a log-normal distribution,
\be
p_{\rm SFMS}(\sfr | \mstar) = \frac{1}{\sqrt{2\pi}\, \sfr \sigma_{\rm SFMS}}\exp\left( -\frac{\log^2 (\sfr / \overline{\psi}_{\rm SFR})}{2 \sigma_{\rm SFMS}^2}\right ).\!\! \label{eq:sfr_sfms}
\ee
There is more freedom in deciding how to model the passive population. The results of \cite{2013MNRAS.431..648W} suggest a relatively uniform distribution of galaxies below the SFMS, while \cite{2013ApJ...767...50M} show a similar log-normal distribution to the SFMS, but shifted to higher $\mstar$ and lower $\sfr$. Semi-analytic models show an even wider range of behaviours, as discussed in \cite{2011MNRAS.416.1566L}. For simplicity, we will use a log-normal distribution of the same form as Eq.~(\ref{eq:sfr_sfms}), but with the mean from Eq.~(\ref{eq:meansfms}) shifted by some factor, and a different scatter,
\bea
\overline{\psi}_{\rm SFR} &\to& a_{\rm pass} \overline{\psi}_{\rm SFR} \nonumber \\
\sigma_{\rm SFMS} &\to& \sigma_{\rm pass}. \label{eq:sfr_pass}
\eea
This more closely follows the observations of \cite{2013ApJ...767...50M}, with the apparent shift in $\mstar$ caused by the transition $f_{\rm passive} \to 1$ as $\mstar$ increases.

\begin{table*}
\renewcommand{\arraystretch}{1.5}
\centering
\begin{tabular}{|l|l|l|r|}
\hline
\multicolumn{4}{|l|}{\bf Stellar mass--halo mass relation, $p(\mstar | \mhalo)$} \\
\hline
$A_\star$ & Overall normalisation & \multirow{4}{*}{Eq.~\ref{eq:msmh}} & 0.018 \\
$\beta_\star$ & Slope at low-mass end & & 1.321 \\
$\gamma_\star$ & Slope at high-mass end & & 0.596 \\
$\log_{10} M^\star_1$ & Mass scale of transition in mean relation ($M_\odot$) & & 12.498 \\ \cline{3-3}
$\sigma_1^\star$ & Overall dispersion amplitude at low masses & \multirow{4}{*}{Eq.~\ref{eq:mstar_sigma}} & 0.557 \\
$\sigma_\infty^\star$ & Dispersion at high-mass end & & 0.031 \\
$\xi_\star$ & Width of transition in dispersion $^\dagger$ & & 4.250 \\
$\log_{10} M^\star_2$ & Mass scale of transition in dispersion ($M_\odot$) $^\dagger$ & & 11.800 \\
\hline
\multicolumn{4}{|l|}{\bf Galaxy type (passive fraction), $f_{\rm pass}(\mstar)$} \\
\hline
$\alpha_f$ & Mass scale of transition (exponent) & \multirow{3}{*}{Eq.~\ref{eq:fpassive}} & 10.804 \\
$\beta_f$ & Determines width of transition & & -2.436 \\
$\zeta_f$ & Determines passive fraction at low mass & & -1.621 \\
\hline
\multicolumn{4}{|l|}{\bf Stellar mass--SFR relation (SFMS galaxies), $p(\sfr | \mstar)$} \\
\hline
$\alpha_{\rm SFMS}$ & Overall normalisation of mean relation (exponent) & \multirow{2}{*}{Eq.~\ref{eq:meansfms}} & -0.077 \\
$\beta_{\rm SFMS}$ & Power-law index of scaling with stellar mass & & 1.037 \\ \cline{3-3}
$\sigma_{\rm SFMS}$ & Dispersion of SFMS & Eq.~\ref{eq:sfr_sfms} & 0.391 \\
\hline
\multicolumn{4}{|l|}{\bf Stellar mass--SFR relation (passive galaxies), $p(\sfr | \mstar)$} \\
\hline
$a_{\rm pass}$ & Determines normalisation of mean relation & \multirow{2}{*}{Eq.~\ref{eq:sfr_pass}} & 0.0011 \\
$\sigma_{\rm pass}$ & Dispersion of passive sequence & & 0.029 \\
\hline
\end{tabular}
 \vspace{1em}
 \caption{Summary of the best-fit parameter values for the main components of the model. The best-fit parameters for the optical magnitude relations are given in Table~\ref{tbl:opticalmag}. Parameters marked with $(\dagger)$ were fixed during the MCMC sampling procedure.}
 \label{tbl:params}
\end{table*}

\subsection{Star-formation -- luminosity relations}
\label{sec:sflum}

Star formation activity releases energy through several processes that cause (typically bright) emission across many bands \citep{1998ARA&A..36..189K}. Young, high-mass stars emit UV radiation, which excites gas in the surrounding inter-stellar medium (ISM). As well as the UV emission itself, this generates line radiation, for example through the de-excitation of hydrogen through H$\alpha$ and H$\beta$ emission. Young stars are short-lived, exploding as supernovae on timescales of order 10 Myr. Supernovae inject high-energy electrons into the ISM, which interact with the galactic magnetic field to emit synchrotron radiation at radio frequencies \citep{1992ARA&A..30..575C}. The process of star formation itself takes place in dense clouds of gas and dust, and the latter, warmed by the collapse process, emits in the infrared.

Star formation-related emission dominates the luminosity of many galaxies. This has led to the development of many ``SFR indicator'' relations \citep{1998ARA&A..36..189K, 2003ApJ...586..794B, 2006ApJ...642..775M} that can be written in the form $L \approx A\, \sfr^{\,\beta}$, with $\beta \approx 1$. In this paper, we will consider only one SFR-related type of emission as an example.
This is continuum radio emission, from synchrotron radiation associated with supernova remnants. Continuum surveys with the SKA and its precursors are expected to detect tens of millions of galaxies from $z \sim 0 - 5$, which can be used for weak lensing and 2D clustering studies \citep{2015aska.confE..18J}. Assuming a powerlaw spectrum for the emission, $S \propto (\nu/\nu_{\rm ref})^\alpha$, with $\alpha \simeq -0.7$, one can use the SFR indicator relation at 1.4 GHz from \cite{2003ApJ...586..794B} to obtain
\be \label{eq:sfradio}
L_{1.4\,{\rm GHz}} = 1.812 \times 10^{28} \left ( \frac{\sfr}{M_\odot {\rm yr}^{-1}} \right ) \,{\rm erg}\,{\rm s}^{-1}\,{\rm Hz}^{-1}. 
\ee
We will omit scatter in this relation, treating the scaling as being deterministic for simplicity. Unlike some other SFR indicators \citep[e.g. H$\alpha$;][]{1998ARA&A..36..189K, 2016arXiv160301453P}, the radio emission does not suffer from extinction, but can be contaminated by (e.g.) free-free and emission from AGN jets. We assume here that free-free is subdominant, and that AGN-dominated and passive galaxies can be selected out, leaving only normal star-forming galaxies.

\subsection{Optical magnitude relations} \label{sec:optlum}

The connection between bulk galaxy properties and emission in optical bands is more convoluted. Optical emission is primarily sourced by the aggregate of all of the stars in the galaxy, which form over an extended period of time, and evolve at different rates depending on their initial mass. The observed optical luminosity is therefore a probe of the star-formation history of the galaxy, not just its present state.

Semi-analytic models typically reconstruct the star-formation history explicitly, evolving an assumed initial mass function forward in time with a stellar population synthesis (SPS) model, and taking into account later bursts of star formation caused by mergers and other events. In the absence of the actual merger history from a simulation, Monte Carlo realisations of plausible merger histories can also be substituted \citep[e.g.][]{2008MNRAS.383..557P}.

A similar approach would be possible here, at the cost of adding significant computational complexity and relying on a `black box' SPS code. Instead, we attempt a simpler approximate treatment to preserve the analytic, parametric nature of the model. We begin by hypothesising that the stellar population can be characterised by two main components: newly-formed stars, dominated by the high-mass end of the IMF; and an older population that formed long before, dominated by lower-mass stars on the main sequence. The former will be bluer, the latter redder.

\begin{figure} 
\centering
\includegraphics[width=0.52\textwidth]{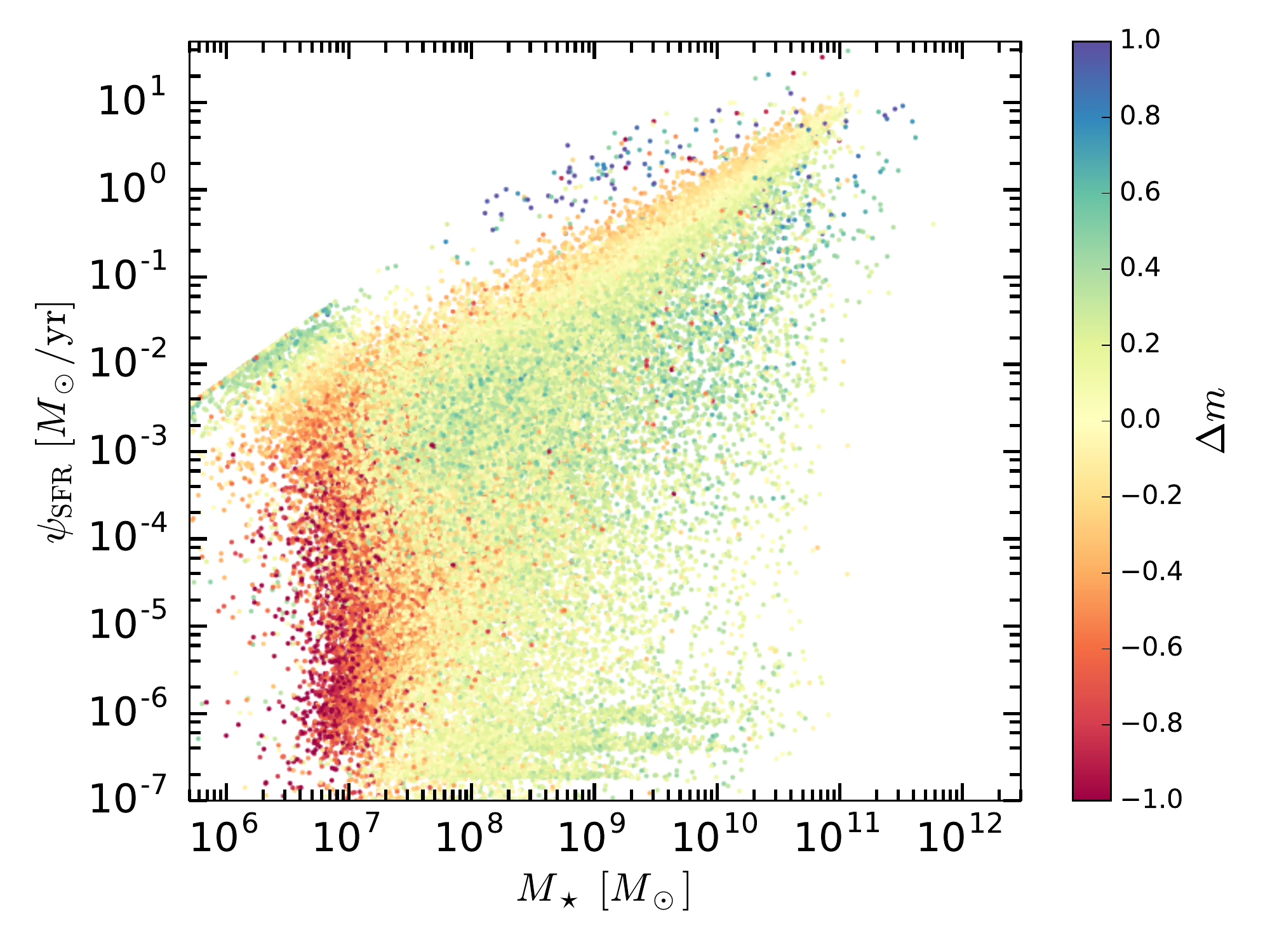} \vspace{-2em}
\caption{Difference between $u$-band magnitudes at $z=0$ (without dust attenuation) in the \citet{Guo:2010ap} simulation, and those predicted by our best-fit relation, Eq.~\ref{eq:mag_fit}, as a function of SFR and stellar mass. Resolution effects in the simulation are apparent at high and low mass, and low SFR.}
\label{fig:opticalresidual}
\end{figure}

Next, we propose an ansatz for the mean absolute magnitude of a galaxy in a given optical band (labelled by $\nu$):
\bea
\overline{m}_\nu(\sfr, \mstar) &=& - c_0^{(\nu)} + A^{\rm opt}_\star \left ( c^{\rm opt}_\star + \left ( \frac{\mstar}{10^9 M_\odot} \right )^{\beta^{\,{\rm opt}}_\star} \right ) \nonumber \\
&& + A^{(\nu)}_\times \left ( \frac{\mstar}{10^9 M_\odot} \right )^{\beta^{\rm opt}_{\times}} \Big ( \sfr \Big )^{\gamma^{\rm opt}_\times}. \label{eq:mag_fit}
\eea
A rough physical interpretation of each term is as follows. The first is an offset that determines a characteristic galaxy magnitude in a given band. The second characterises the total emission from the older stellar population, using stellar mass as a proxy. The final term represents the contribution from the younger population, with SFR as a proxy. Since there is a relationship between SFR and stellar mass, this term must also include an $\mstar$-dependent factor to model the `mixing' between the two. The amplitude of the final term also depends on the band; star formation contributes less to the total luminosity in redder optical bands.

To calibrate the relation, we use the public semi-analytic catalogues of \cite{Guo:2010ap}, which provide $ugriz$ absolute magnitudes with and without a dust extinction correction. The residuals of the best fit to the unattenuated $u$-band magnitudes at $z=0$ are shown in Fig.~\ref{fig:opticalresidual}, as a function of $\mstar$ and $\sfr$. The distribution of residuals is reasonably simple, with a mean close to zero and only a low level of structure present. The standard deviation of the residuals across the whole plane is $\sigma(\Delta m_\nu) \sim 0.3$ mag in all 5 bands, which is quite small considering the simplicity of this approach; it has only five global parameters and two free parameters per band. Ignoring the dependence of the residuals on $\mstar$ and $\sfr$, we find that they are well modelled by a shifted log-normal distribution,
\be
p(m_\nu | \,\sfr, \mstar, z) = \frac{1}{\sqrt{2\pi} x\, \sigma^{(\nu)}_{m}} \exp \left ( -\frac{1}{2} \left (\frac{\log x - \mu^{(\nu)}}{\sigma^{(\nu)}_{m}} \right )^2 \right ), \label{eq:opticalpdf}
\ee
where $x \equiv \overline{m}_\nu - m_\nu + b_\nu,\, x > 0$. Here, $\log b_\nu \equiv {\,\mu^{(\nu)} + (\sigma^{(\nu)}_m)^2/2}$, where $b_\nu$ is the mean of the shifted log-normal distribution and $\mu^{(\nu)}$ and $\sigma^{(\nu)}_m$ are free parameters in each band. These parameters are calibrated by performing a least-squares fit of a log-normal pdf to the histogram of residuals, independent of $\sfr$ and $\mstar$.

As shown in Fig.~\ref{fig:opticalresidual}, the residuals do have some dependence on $\sfr$ and $\mstar$. The residuals can also be correlated between bands. While we have neglected structure like this in our model as a simplifying assumption, such correlations are partially responsible for the observational patterns seen in `colour-magnitude' diagrams, which are often used to select different populations of galaxies. In its current state, the model will therefore be unable to reproduce features such as the red sequence or blue cloud \citep[e.g.][]{2001AJ....122.1861S} for example.

\subsection{Optical dust extinction} \label{sec:opticalext}

Some fraction of the optical emission from galaxies is absorbed by the dust that they contain. The amount of absorption depends on a number of factors, including wavelength (bluer colours are more strongly absorbed); morphology and inclination angle (light from the bulge and disk components is affected differently); and the dust content and its distribution inside the galaxy. Intrinsic dust absorption is difficult to simulate or characterise with observations, which is problematic -- accounting for dust is a necessary step in connecting optical emission to the physical properties of galaxies, as well as being important for modelling infrared luminosities.

\begin{figure} 
\includegraphics[width=0.48\textwidth]{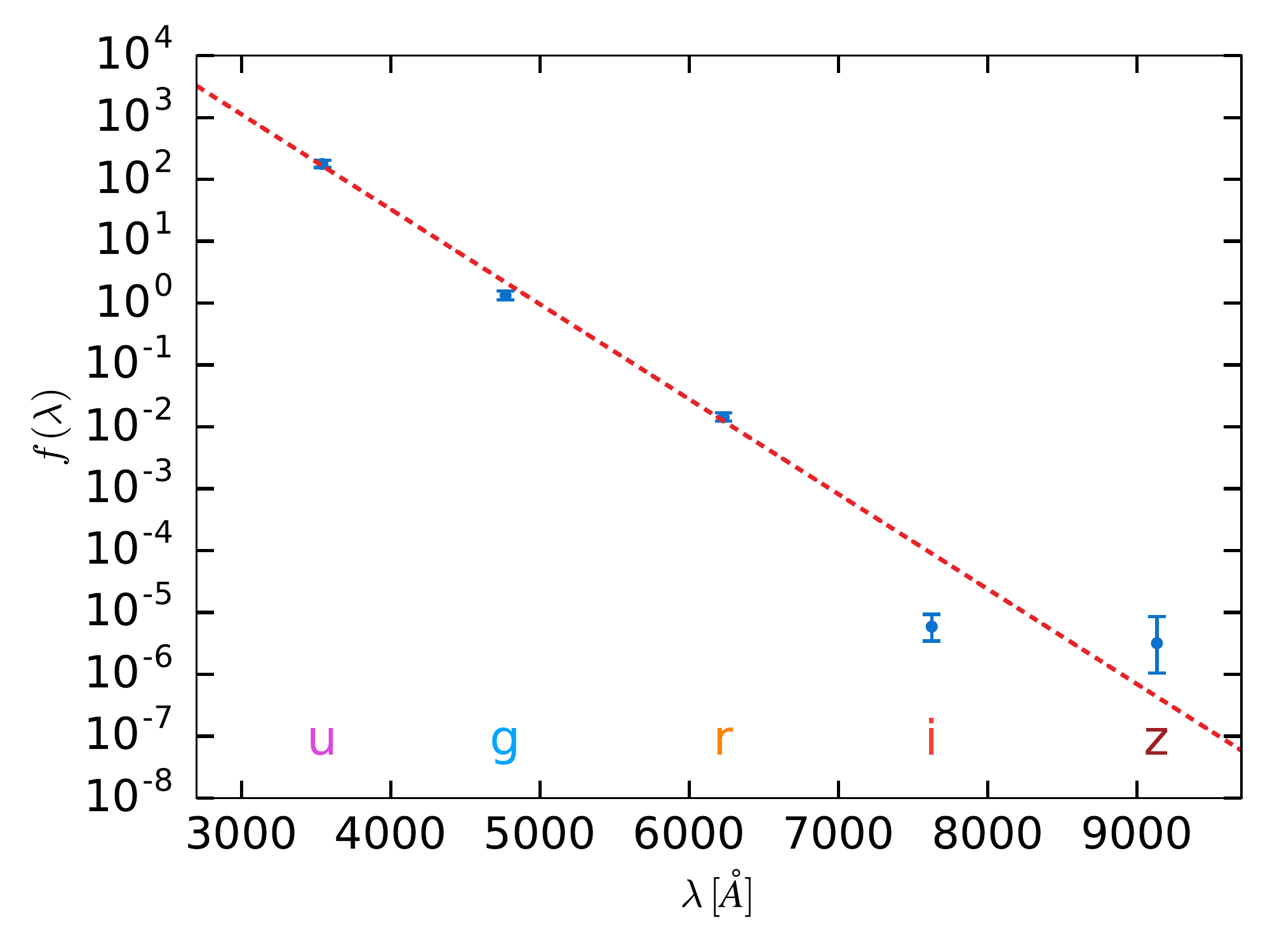}
\caption{Amplitude of the optical dust attenuation as a function of wavelength, $f(\lambda)$. The blue datapoints show the median and 68\% credible intervals for $f(\lambda_j)$ in each band $j$, after conditioning on the best-fit values of all other parameters. The red dashed line shows the relation from Eq.~\ref{eq:flambda} for the best-fit values of $\tau_0$, $\kappa$, and $\lambda_0$.}
\label{fig:attenfreq}
\end{figure}

There have been numerous attempts to construct attenuation models from observations and simple considerations such as galaxy morphological properties \citep[e.g.][]{2004A&A...419..821T, 2007MNRAS.379.1022D}. Several dust models that have been used in semi-analytic simulations are compared in \citet{2009MNRAS.392..553F}. Many such models use an empirically-determined spectral dependence from \citet{2000ApJ...533..682C}, which has been found to offer a good description of the corrections required to get SFR determinations at different wavelengths to agree \citep[e.g.][]{2009ApJ...698L.116P}.

In keeping with our analytic approach, we will adopt a simpler parametric model. An ansatz for the optical depth due to internal dust extinction is
\be \label{eq:tau}
\tau(\lambda) = f(\lambda)\, \left ( \frac{\mstar}{10^{10} M_\odot} \right )^{\beta_\tau} \left ( 1 + a_{\rm disk} \sin \theta \right ),
\ee
where $f(\lambda)$ is the attenuation amplitude as a function of (rest-frame) wavelength, $a_{\rm disk}$ is the effective attenuation due to the disk, and $\theta$ is the inclination angle of the disk with respect to the line of sight. The frequency dependence (see Fig.~\ref{fig:attenfreq}) can be modelled approximately as
\be
f(\lambda) = \tau_0 \, \exp \left [{-\kappa (\lambda - \lambda_0)} \right ], \label{eq:flambda}
\ee
where $\tau_0$ is an overall normalisation parameter, $\kappa$ determines the scaling with wavelength, and $\lambda_0$ is a reference wavelength. The change in magnitude due to the dust attenuation is
\be
\Delta m \equiv m_{\rm obs} - m_{\rm int} = 5 \log_{10} \left ( \frac{S_{\rm int}\,e^{-\tau}}{S_{\rm int}} \right ) = 1.086\, \tau,
\ee
where $S_{\rm int}$ is the intrinsic flux before attenuation.

\begin{table}
\renewcommand{\arraystretch}{1.5}
\centering
\begin{tabular}{lrrrrr}
\hline
{\bf Band:} & \multicolumn{1}{|c|}{\bf u} & \multicolumn{1}{|c|}{\bf g} & \multicolumn{1}{|c|}{\bf r} & \multicolumn{1}{|c|}{\bf i} & \multicolumn{1}{|c|}{\bf z} \\
\hline
$A_\star$ & \multicolumn{5}{|c|}{3876.6} \\
$\beta_\star$ & \multicolumn{5}{|c|}{-0.000243} \\
$c_\star$ & \multicolumn{5}{|c|}{-1.011} \\
\hline
$A_\times^{(\nu)}$ & -3.612 & -2.700 & -2.016 & -1.726 & -1.564 \\
$\beta_\times$ & \multicolumn{5}{|c|}{-0.263} \\
$\gamma_\times$ & \multicolumn{5}{|c|}{0.290} \\
\hline
$c_0^{(\nu)}$ & -27.318 & -25.945 & -25.284 & -24.982 & -24.810 \\
$\mu^{(\nu)}$ & -0.066 & -0.054 & -0.019 & -0.009 & -0.009 \\
$\sigma_m^{(\nu)}$ & 0.281 & 0.252 & 0.247 & 0.253 & 0.272 \\
\hline
$\tau_0$ & \multicolumn{5}{|c|}{1.041} \\
$\lambda_0$ & \multicolumn{5}{|c|}{4977.2} \\
$\kappa$ & \multicolumn{5}{|c|}{0.004} \\
$a_{\rm disk}$ & \multicolumn{5}{|c|}{1.429} \\
$\beta_\tau$ & \multicolumn{5}{|c|}{5.478} \\
\hline
\end{tabular}
  \caption{Best-fit values for the parameters of the optical magnitude model, including dust attenuation. The non-attenuation parameters were fitted to the $z = 0$ simulated galaxy catalogue of \citet{Guo:2010ap}. The attenuation parameters $a_{\rm disk}$ and $\beta_\tau$ were fitted to the $g$ and $z$-band GAMA LF data simultaneously with the non-optical model parameters (see below). The attenuation parameters $\tau_0$, $\lambda_0$, and $\kappa$ were fitted using a maximum likelihood method on the full 5-band GAMA LF data (conditioned on all other model parameters).}
  \label{tbl:opticalmag}
\end{table}

The two factors in parentheses in Eq.~(\ref{eq:tau}) are motivated as follows. First, the amount of dust in the galaxy is assumed to scale with the stellar mass to some power. Dust builds up in the interstellar medium as successive populations of stars evolve and die. While star formation is associated with thermal dust emission, this only represents one component of the total dust content of a galaxy; stellar mass should be more indicative of the integrated star formation history that has actually generated the dust. There is observational support for a relatively steep power-law scaling of the dust attenuation with stellar mass \citep[e.g.][]{2009ApJ...698L.116P, 2012A&A...545A.141B, 2014MNRAS.437.1268H}, but we will not enter into a discussion here.

Secondly, extinction will be greater when the disk is seen edge-on, i.e. when $\theta = \pi/2$. The relative optical depth of the disk (compared to the bulge) is represented by a scaling factor, $a_{\rm disk}$. For random galaxy orientations, $\sin \theta$ follows a uniform distribution. If we assume that inclination angle is the only random variable in the optical depth expression, it follows that
\be
p(m_\nu^{\rm obs} | m_\nu^{\rm int}) = {\rm Uniform}[m_\nu^{\rm int} + \Delta m(\theta=0), m_\nu^{\rm int} + \Delta m(\theta=\pi/2)].\nonumber
\ee
In the absence of explicit modelling of other stochastic contributions to the optical depth, it is likely that the inclination angle term will absorb other sources of scatter in the observations. As such, the $a_{\rm disk}$ parameter should not be interpreted directly as a physical quantity.

Since one only ever observes the dust-attenuated flux in a given band, it is useful to immediately marginalise out the intrinsic magnitude,
\be
p(m_\nu^{\rm obs} | \mstar, \sfr) = \int p(m_\nu^{\rm obs} |\, m_\nu^{\rm int}) p(m_\nu^{\rm int} | \mstar, \sfr)\, dm_\nu^{\rm int}.
\ee
The integral can be evaluated analytically for our model pdfs;
\bea
p(m_\nu^{\rm obs} | \mstar, \sfr) = \frac{1}{2} \frac{{\rm erf}\Big(y(m_\nu^{\rm obs}, \pi/2)\Big) - {\rm erf}\Big(y(m_\nu^{\rm obs}, 0)\Big)}{\Delta m_\nu(\pi/2) - \Delta m_\nu(0)},
\eea
where the denominator comes from the normalisation of the uniform distribution, and the argument of the error functions is
\bea
y(m_\nu^{\rm obs}, \theta) &=& \frac{\log ( x_\nu(\theta)) - \mu^{(\nu)}}{\sqrt{2}\, \sigma_m^{(\nu)}}, \\
x_\nu(\theta) &=& {\rm max} \left\{\, \overline{m}^{\,\rm int}_\nu + \Delta m_\nu(\theta) + b_\nu - m_\nu^{\rm obs}, ~~ 0 \,\right \}.
\eea
The dependence on $\mstar$ and $\sfr$ enters through $\overline{m}^{\,\rm int}_\nu$ and $\Delta m_\nu$.

It should be noted that this is a rather basic statistical model for intrinsic dust extinction. We have assumed that the optical depth in all galaxies follows the same mean relation, and (effectively) differs only in inclination angle. The other parameters of the model will surely also vary from galaxy to galaxy though, especially depending on its morphology -- passive elliptical galaxies do not have separate bulge and disk components, for example. A simple extension to the model would be to choose $a_{\rm disk}$ separately for star-forming and passive galaxies, but we do not pursue this here. As such, this aspect of the model should be considered preliminary -- it is able to fit the observations, as we shall see in the next section, but any physical interpretation of the fits should proceed with caution.

\section{Model calibration and parameter dependence} \label{sec:params}

A set of parameter values is needed as an input to the model. While best-fit parameters for most of the conditional distributions in Fig.~\ref{fig:model} are available separately in the literature, suitable parameter values can also be obtained by fitting the entire model to a set of diverse observational (or simulated) data. This has the advantage of ensuring consistency between the various components of the model. We will pursue the latter approach here, using a Monte Carlo sampling method to derive the joint posterior distribution of the model parameters given several datasets from galaxy surveys at different wavelengths. 

In the first part of this section, we describe the data used to perform the fits: recent constraints on the radio luminosity function of star-forming galaxies, and the luminosity function of optically-detected galaxies, all at $z \approx 0$. We also describe the sampling method, and cuts that were used on the data. We then report the parameter constraints that were obtained, and analyse them for consistency using simple goodness-of-fit statistics. We also compare with the best-fit values that were found for several of the model components independently by previous studies.

\subsection{Input data}

We use the following data to constrain the model:
\begin{itemize}
 \item The luminosity function of star-forming radio galaxies cross-identified in NVSS and 6dFGS \citep{2007MNRAS.375..931M}. This mostly constrains the SFMS and halo mass-stellar mass pdfs, and does not require a dust attenuation correction.
 \item The $ugriz$ optical luminosity functions from GAMA \citep{2012MNRAS.420.1239L}. These are corrected for extinction in the Milky Way, but not for the internal extinction in the target galaxies. We fit both the $g$ and $z$-band data to the full model; the $z$ band is least affected by extinction, while the $g$ band is strongly affected, allowing some of the parameters of the dust attenuation model to be constrained. Data from the other bands are not used in the general fits, but they are used to constrain the dust extinction amplitude parameter (conditioned on the best-fit parameters of the general fit; see below).
 \item The joint distribution of stellar mass, star-formation rate, and intrinsic (zero extinction) optical magnitudes from the \cite{Guo:2010ap} semi-analytic simulation. This is used to calibrate the optical magnitude conditional distributions only, as described in Sect.~\ref{sec:optlum}. The resulting best-fit parameters are then fixed throughout the rest of the analysis.
\end{itemize}
These datasets were chosen because they are sufficient to constrain all of the key parameters of the model at $z \approx 0$. Other datasets or simulations could also have been included, but this is beyond the scope of the present work. Note that most of these datasets assume different background cosmologies; we applied corrections to convert them to our fiducial cosmology as appropriate.

\begin{figure*} 
\hspace{-2.5em}
\includegraphics[width=1.03\textwidth]{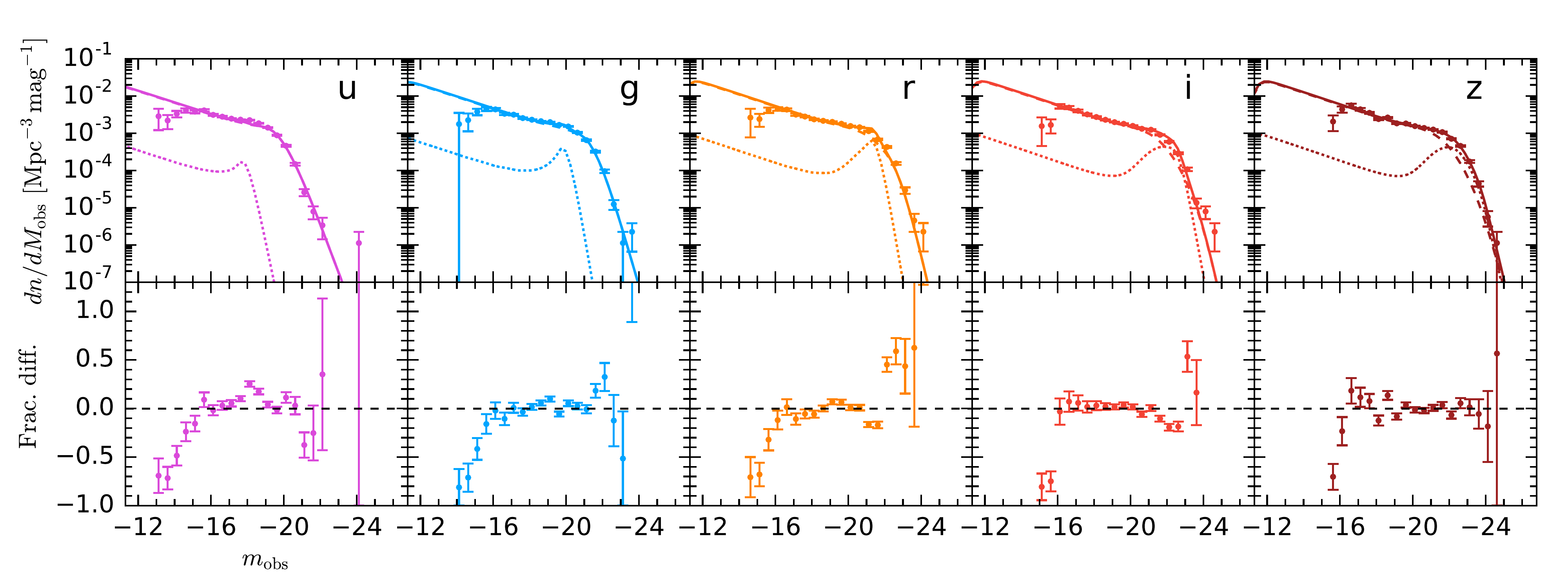}
\caption{{\it Upper panels:} Optical luminosity functions in the $ugriz$ bands at $z \simeq 0$, fitted to data from GAMA (points with errorbars) and the NVSS radio luminosity function. Only the $g$ and $z$ bands were used in the fitting procedure. The solid coloured lines show the total luminosity functions, including the dust attenuation correction; dashed lines show the LF from the star-forming main sequence only; and dotted lines show the passive galaxy LF. {\it Lower panels:} Fractional difference between the best-fit optical luminosity function from our model and the GAMA datapoints.}
\label{fig:lfoptical}
\end{figure*}

We construct independent, approximate likelihoods for each dataset, based on the binned data reported in the literature. The binned GAMA luminosity functions are reported with symmetric error bars. We assume that these data are Gaussian distributed, and that the errors are independent, yielding a likelihood of the form
\be
\log \mathcal{L}(\theta) = -\frac{1}{2} \sum_i \left ( \frac{\Phi_i - \Phi(\theta)}{\sigma_i} \right )^2,
\ee
where $i$ labels the magnitude bins of the luminosity function, $\Phi = dn/dm$, and $\theta$ denotes the set of parameters for which the model luminosity function is evaluated. The optical luminosity function is calculated from the model as
\bea
\Phi(m_\nu^{\rm obs}) \equiv \frac{dn~~}{dm_\nu^{\rm obs}} &=& \int d\sfr \int d\mstar \int d\log\mhalo \nonumber\\
&&~~~~~~ \times\,\, n(\mhalo)\, p(\mstar | \mhalo)\, p(\sfr | \mstar) \nonumber\\
&&~~~~~~ \times\, p(m_\nu^{\rm obs} |\, \sfr, \mhalo).
\eea

The NVSS/6dFGS radio luminosities are reported with asymmetric errorbars. We were unable to find published posterior distributions for the data, and so we use an approximate likelihood for data with asymmetric errors from an unknown distribution, taken from \cite{2003physics...6138B}:
\bea
&&\log \mathcal{L}(\theta) \approx -\frac{1}{2} \sum_i x_i^2(\theta) \left ( 1 - 2A_i x_i(\theta) + 5 A_i^2 x_i^2(\theta)\right );~~~~~~~~~~~~~~~ \\
&& x_i(\theta) = \frac{y_i - \hat{y}_i(\theta) }{\sigma_i}; ~~~
\sigma = \frac{\sigma_+ + \sigma_-}{2}; ~~~
A = \frac{\sigma_+ - \sigma_-}{\sigma_+ + \sigma_-},~~~~~~~~~~~~~~~
\eea
where $y_i$ is the observed value, $\sigma_\pm$ are the upper and lower asymmetric errorbars on $y_i$, and $\hat{y}_i(\theta)$ is the model value for a given set of parameters. This limits to a Gaussian likelihood for symmetric error bars. Since the studies in question report asymmetric errors on the logarithm of the radio luminosity function, we will work directly with $y_i = \log \Phi_i$ as the random variate. Again, this is an approximate form for the likelihood {\it in the absence of more information} about the posterior distributions. If we had access to the actual posteriors, this treatment would be unnecessary. The radio luminosity function is calculated in our model as
\bea
\Phi(L_{\rm rad}) \equiv \frac{dn}{d\log L} &=& \int d\sfr \int d\mstar \int d\log\mhalo \nonumber\\
&&~~~~~~ \times\,\, n(\mhalo)\, p(\mstar | \mhalo)\, p_{\rm SFMS}(\sfr | \mstar) \nonumber\\
&&~~~~~~ \times\, L_{\rm rad} \, p(L_{\rm rad} |\, \sfr),
\eea
where only the SFMS is used; passive galaxies are assumed to have been selected out.

The final log-likelihood is simply the sum of the log-likelihoods for each dataset, i.e. two optical bands and the radio.

\subsection{Sampling method}

We use the {\tt emcee} affine-invariant Markov Chain Monte Carlo (MCMC) sampler \citep{2013PASP..125..306F} to sample from the posterior distribution. We run this for 2000 samples for each of 128 workers, leaving 5500 samples after discarding burn-in and thinning the chains.

Seventeen (17) parameters were sampled by the MCMC (see Table~\ref{tbl:params} for a summary). We did not sample the parameters of the optical luminosity relation defined in Sect.~\ref{sec:optlum}, as this would allow too much freedom in the model. The posterior distributions of other parameters could shift if these were allowed to vary, although this would likely also lead to degeneracies. We did sample three of the dust attenuation parameters however (see Table~\ref{tbl:opticalmag}). The starting position of the chains was chosen based on a rough visual fit to the input data, which aimed to find parameter values broadly similar to the best-fit values from the literature. This procedure was necessary to help avoid local maxima of the likelihood.

Simple prior bounds were chosen to keep the walkers in a physically reasonable region of parameter space. Certain features can be reproduced by allowing the SFMS to be too broad or too narrow for example, and the passive and star-forming sequences can switch position in the SFR-$\mstar$ plane if the amplitude parameters are not kept in check. Negative dispersion parameters should also be disallowed, as otherwise the pdfs become ill-defined. To avoid problems like these, we chose the following priors:
\bea
0.05 \le& \sigma_{\rm SFMS} &\le 1 \\
0.01 \le& \sigma_{\rm pass} &\le 2 \\
0.02 \le & \sigma_1^\star,~~ \sigma_\infty^\star & \\
0.001 \le & a_{\rm pass} &\le 0.9 \\
11.6 \le& \log_{10} M_1^\star &\le 12.5.
\eea
The last prior in this set was chosen to restrict the transition scale of the stellar mass-halo mass relation to be the same order of magnitude as the one found by \citet{2010ApJ...710..903M}. Allowing this to be completely free often resulted in the walkers finding local minima with substantially different best-fit parameters from others in the literature.

We identified the burn-in period by plotting the mean of the log-likelihood over all 128 workers as a function of sample number. The mean log-likelihood stops improving as the chains enter the maximum likelihood region, which was seen as a flattening of the plot after a certain number of steps. To thin the chains, we calculated the integrated autocorrelation time \citep{sokal1997monte} for the chain from each worker as a function of thinning factor,
\be
\tau_{\rm int}(f) = \frac{1}{2} \sum_\tau \rho(\tau, f),
\ee
where $\rho(\tau, f)$ is the normalised autocorrelation function at lag $\tau$ and $f$ is the thinning factor. We chose the smallest thinning factor that reduced the autocorrelation time below unity, and then combined the thinned chains from all workers into a single chain. As jumps between workers are allowed by ensemble samplers, the chains from each worker are correlated with one another, so this procedure does not fully decorrelate the samples. We verified that the sample means and standard deviations for a few parameters did not shift significantly when different thinning factors or burn-in periods were used though, or when different numbers of workers were kept in the final chains. The posterior distribution for all 17 parameters is shown in Fig.~\ref{fig:triangle}, and the processed chain is available to download from \url{http://www.philbull.com/ghost}.

Once a best-fit model has been obtained from the MCMC, we then fit the frequency-dependent dust attenuation parameters $\tau_0$, $\kappa$, and $\lambda_0$ (see Eq.~\ref{eq:flambda}) to all five optical bands simultaneously, using a maximum likelihood method. The likelihood for the optical luminosity functions is evaluated by varying these parameters while keeping all other parameters fixed to their best-fit values. The results of this procedure are shown in Table~\ref{tbl:opticalmag} and Fig.~\ref{fig:attenfreq}.

\begin{figure} 
\includegraphics[width=0.48\textwidth]{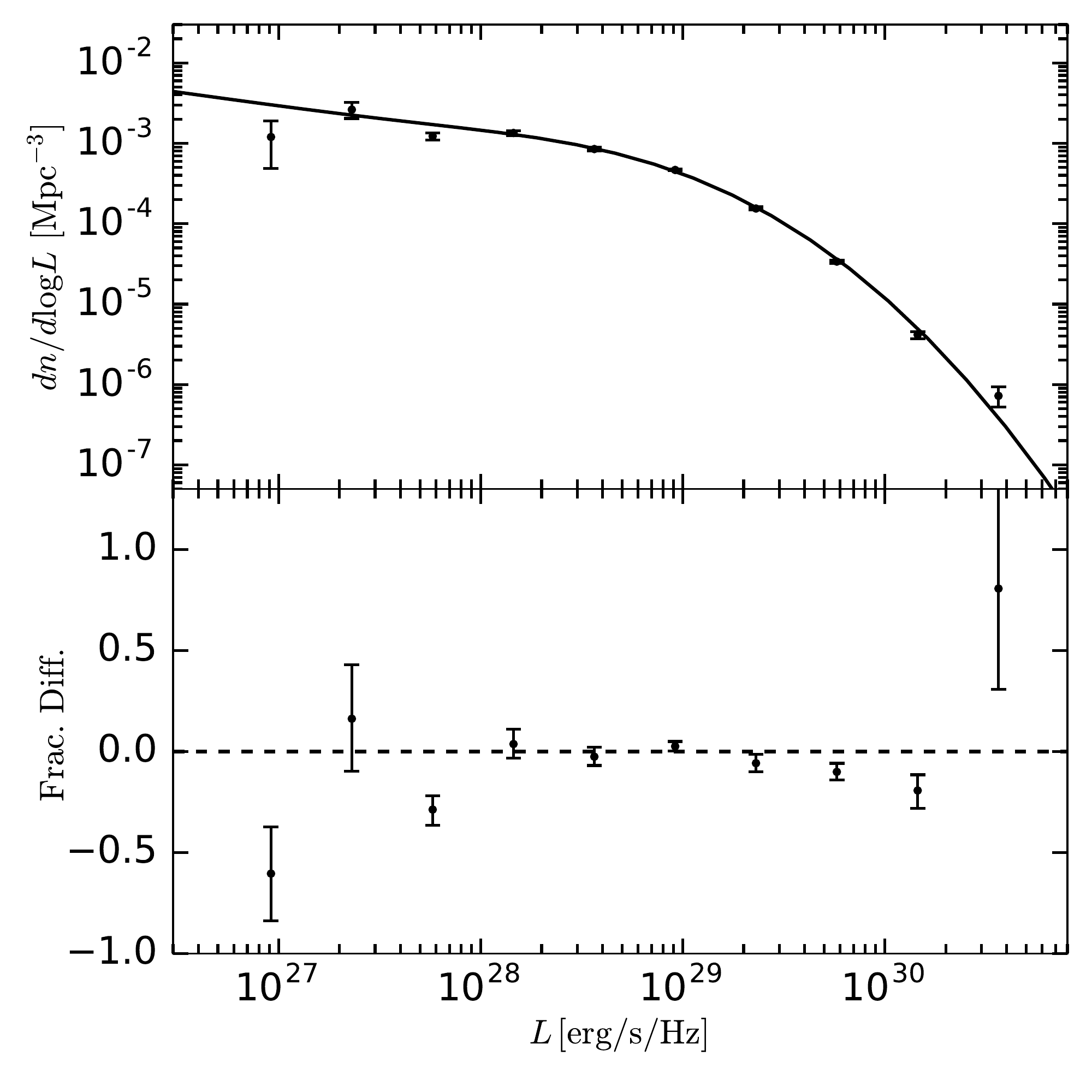}\vspace{-2em}
\caption{{\it Upper panel:} Luminosity function of star-forming radio galaxies from the best-fit model (solid black line) to the GAMA and NVSS data, with the latter data shown as black points. {\it Lower panel:} The fractional difference between the best-fit model and the NVSS datapoints.}\vspace{-1em}
\label{fig:lfradio}
\end{figure}

\subsection{Analysis of the model fits} \label{sec:analysis}

The MCMC procedure described above results in a set of samples from the posterior distribution of the model and chosen data. The parameter values of the best-fit (i.e. maximum likelihood) model from the chains are summarised in Table~\ref{tbl:params}. In this section, we provide a brief analysis of how well the global best-fit model fits the input data: the optical and radio luminosity functions from GAMA and NVSS respectively. (The fit to the simulated optical magnitude data was discussed in Sect.~\ref{sec:opticalext}.)

Fig.~\ref{fig:lfradio} shows the luminosity function of star-forming radio galaxies at $z \simeq 0$ from the global best-fit model, along with the NVSS data. The residuals are relatively small, with no large deviations across the full luminosity range. The model has a $\chi^2$ of $30.1$ for 10 datapoints however, which is formally a poor fit (although recall that we used an approximate, non-Gaussian likelihood function, and have assumed that the errors are independent). This must be compared to the performance of other methods to model luminosity functions though, which also tend to produce formally poor fits. For example, if we fit the phenomenological Schechter-like luminosity function suggested by \citet{2007MNRAS.375..931M} to the radio LF only, using the same likelihood function, we obtain a best-fit model with $\chi^2 = 18.6$ (for 4 free parameters), which gives a probability-to-exceed of 0.005 -- a better, but still poor fit. We conclude that our model fit is acceptable for simple modelling purposes, due to the lack of significant deviations, but that it does not provide a complete, statistically acceptable description of the data.

The best-fit model for the optical LFs is shown in Fig.~\ref{fig:lfoptical}. The residuals are again reasonably small, except for noticeable deviations at the extreme bright and faint ends in most bands, and a bump feature at $m_{\rm obs} \approx -18$ in the $u$-band. The goodness-of-fit for the bands that were used in the MCMC were again relatively acceptable but formally poor, with $\chi^2 = 39.5$ (49.2) and $34.1$ (34.1) for the $g$ and $z$ bands, over 17 and 16 datapoints respectively (where the numbers in parentheses denote the $\chi^2$ after the frequency-dependent dust attenuation parameters have been fitted to all five bands). The $u$, $r$, and $i$ bands, which were not included in the MCMC, have $\chi^2 = 195.2, 156.7, 87.8$ (161.0, 156.0, 85.0) respectively, with 17 datapoints each. 

The deviations at the faint end of the LFs in Fig.~\ref{fig:lfoptical} are likely due to completeness effects, where galaxies have gone undetected near the flux limit of the survey. While a completeness correction was applied to the data by \citet{2012MNRAS.420.1239L}, a systematic drop remains in the last few datapoints in all bands, which indicates a residual completeness systematic. We discarded the last 3 points in each band in the MCMC to avoid this skewing our results.

The deviations at the bright end are less severe, due to the larger errorbars there. These could be caused by a model error; we do not explicitly include a bright starburst galaxy population for example, but this would contribute most significantly at the bright end. The deviation could also be explained by selection effects, e.g. due to bright galaxies being intrinsically rarer and thus more likely to be under-sampled in surveys that cover limited volumes (as is always the case at $z \simeq 0$). Intrinsic dust attenuation also affects bright galaxies most significantly, so a model error or a selection effect related to this could also cause a discrepancy.

\begin{figure} 
\includegraphics[width=0.48\textwidth]{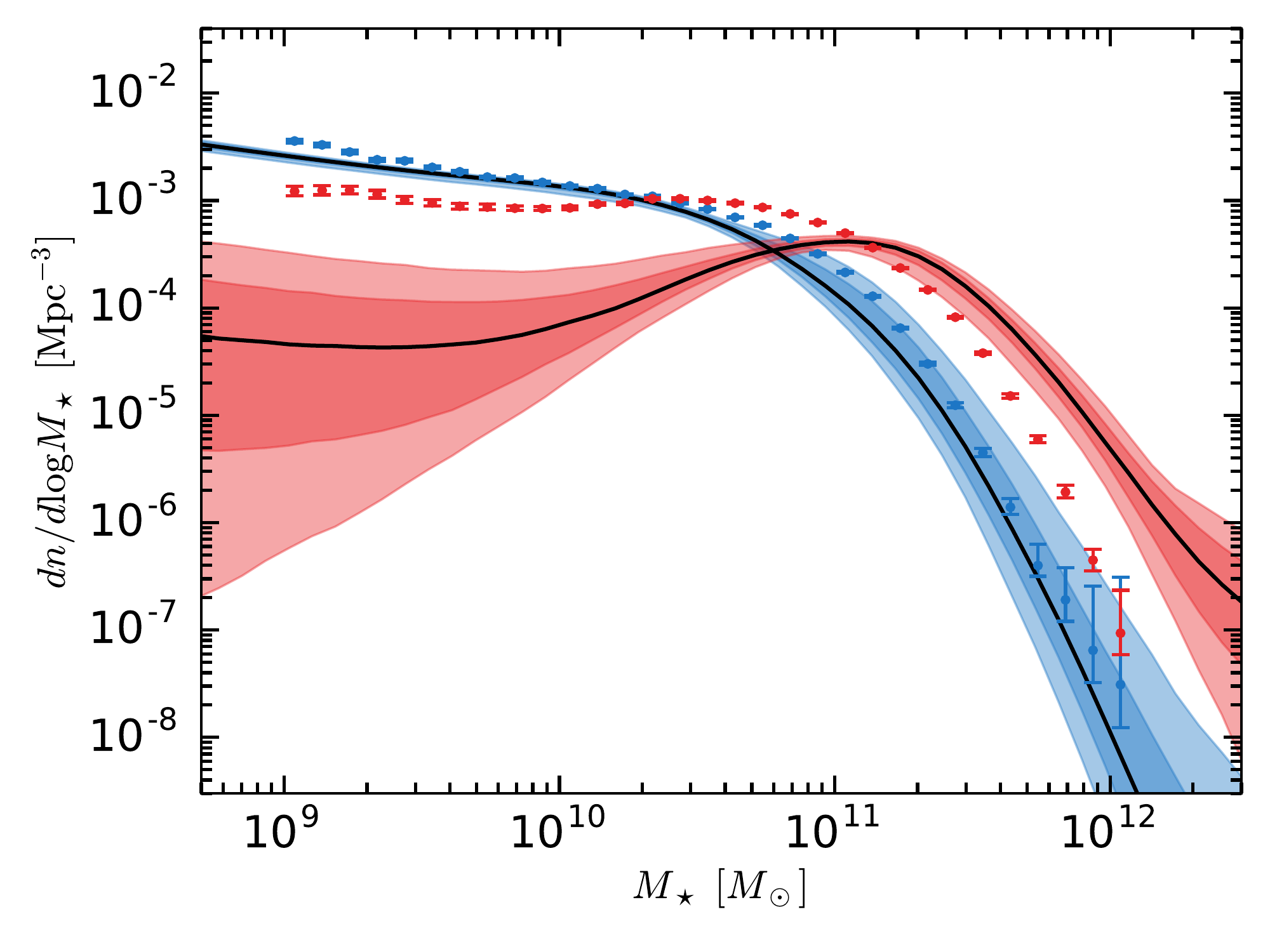} \vspace{-1em}
\caption{Stellar mass function for SFMS (blue) and passive (red) galaxies. Datapoints are from SDSS/GALEX at $z \approx 0.1$ \citep{2013ApJ...767...50M}. Solid black lines with coloured bands show the predictions of our global best-fit model, including 68\% and 95\% credible regions from the MCMC. Our model fits did not use the SDSS/GALEX data, or any other SMF data; these curves are predictions.}\vspace{-1em}
\label{fig:smf}
\end{figure}

The bump feature in the $u$-band is significant, even if the fit is otherwise (qualitatively) a good one. There is a slight hint of such a feature at around the same position (i.e. a little fainter than the faint-end transition in the LF) in all of the bands. The best-fit model tries to explain this by allowing the LF of the sub-dominant passive sequence to form a relatively sharp peak at around this magnitude. This is achieved by driving the width of the passive sequence, $\sigma_{\rm pass}$, to low values. Assuming that this bump is a real feature of the data, the failure of our $u$-band LF to reproduce it could be due to the mean optical magnitude model, Eq.~\ref{eq:mag_fit}. A systematic trend in the width and height of the bump in the passive LF can be seen in Fig.~\ref{fig:lfoptical}; both decrease towards bluer wavelengths. The data could likely be explained better if the bump was higher in the $u$-band, which may be achievable with a slightly different set of optical magnitude relation parameters for this band, or by taking into account the structure of the residuals in the $\sfr-\mstar$ planes (see Fig.~\ref{fig:opticalresidual}).

\begin{figure} 
\includegraphics[width=0.48\textwidth]{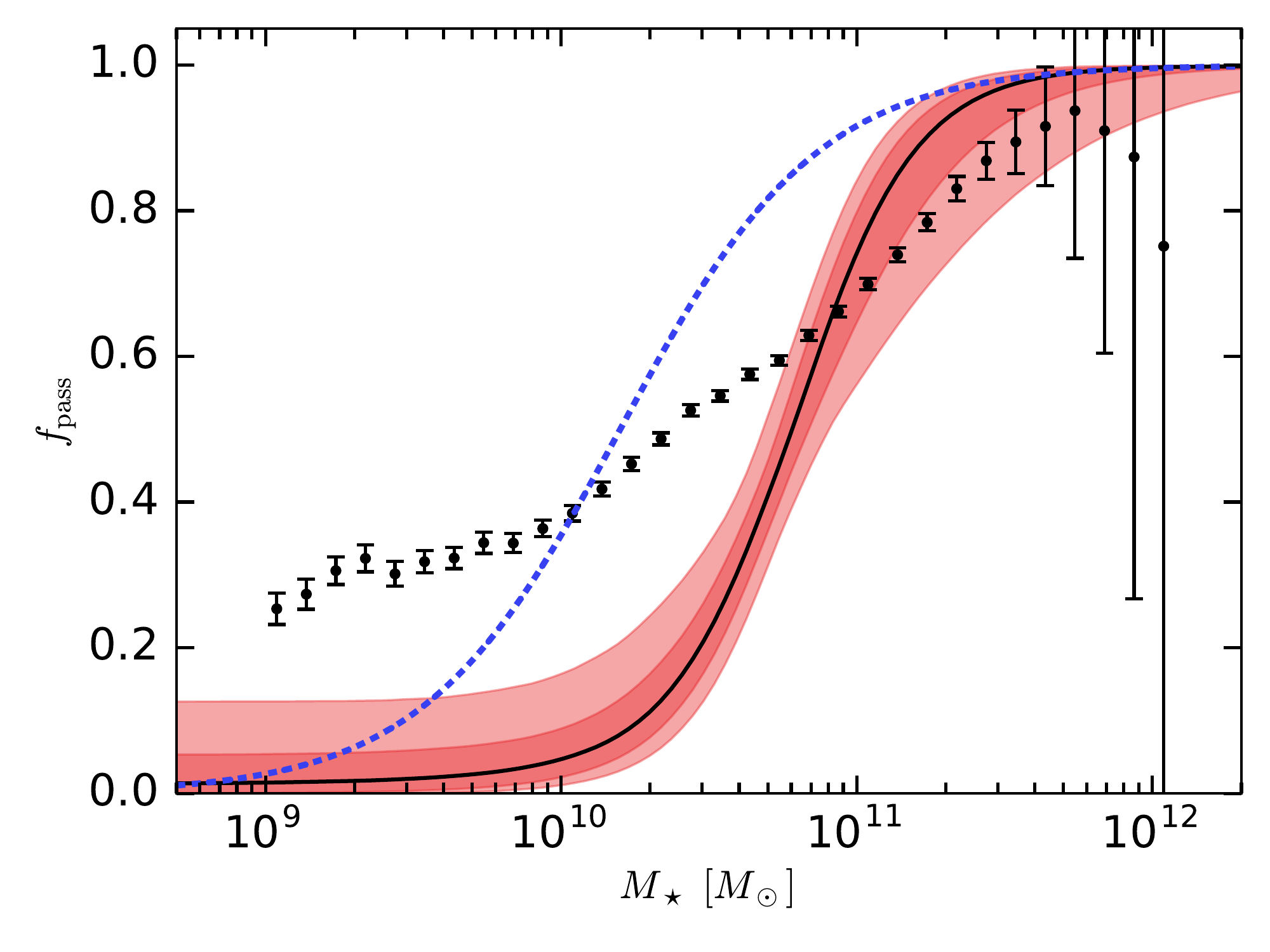}\vspace{-1em}
\caption{The predicted fraction of passive galaxies as a function of stellar mass. Red bands show the 68\% and 95\% credible intervals from our model, when fitted to the GAMA and NVSS optical/radio luminosity functions. The purple dashed line shows the simpler relation assumed in \citet{2013ApJ...770...57B}. The black datapoints show the passive fraction from SDSS/GALEX, derived by dividing the reported quiescent stellar mass function by the sum of the quiescent and star-forming SMFs, and then estimating the errors using simple Gaussian error propagation.}\vspace{-1em}
\label{fig:fpass}
\end{figure}

For the dataset that was directly fitted for (i.e. the $g$ and $z$ band optical LFs and radio LF), we found an effective $\chi^2$ of $103.7$ for 43 datapoints and 17 free parameters, giving roughly $26$ degrees of freedom when the datapoints are assumed independent. For the most optimistic hypothesis, that the data and errorbars are correctly estimated and free of systematic errors, and that our model is the correct underlying description of the data, the probability to exceed this $\chi^2$ is negligible ($p \lesssim 3\times 10^{-11}$), implying that this hypothesis is incorrect. For the full dataset, including the other three optical bands and the maximum likelihood-fitted frequency-dependent dust attenuation parameters, the total $\chi^2$ is 515.5 over (effectively) 19 free parameters and 94 datapoints.

Our overall conclusion is that the data are qualitatively well-fit by the model, which produces luminosity functions that closely follow the shape and amplitude of the LF data. If one assumes that the errors on the data are independent, the goodness-of-fit is technically poor however. Numerous effects could contribute to cause this, one of which is likely to be the fact that the model is rather simple. Nevertheless, the results we obtained will be sufficient for many modelling purposes.

\subsection{Consistency with previous results}

In this section, we compare our global best-fit model with a small selection of constraints on individual components of the model from the literature. Since we use only the radio and optical LFs, and semi-analytic model fits to the optical magnitude relation, to constrain the model, the constraints we place on other components are indirect. Comparing with direct constraints should then give us a good idea of the model's consistency.

Fig.~\ref{fig:smf} shows the stellar mass function for the SFMS and passive populations for the global best-fit model, compared with data from SDSS/GALEX at $z \approx 0.1$ \citep{2013ApJ...767...50M}. Neither the SFMS or passive SMF model predictions are a good fit to the data. The SFMS stellar mass function has a similar shape to the data, and fits well at $\mstar \sim 10^{10} M_\odot$, which is encouraging given that this is a prediction based on a fit to very different data. The amplitude of the high-mass end is systematically low however, by 2-3 standard deviations per datapoint, which accumulates to yield a poor overall goodness of fit. The passive SMF is a much worse fit, despite the relatively large uncertainty at high and low stellar mass from the MCMC.

This discrepancy could be caused by a number of different effects. For example, the categorisation into star-forming and passive galaxies assumed by our model could differ markedly from the selection used by \citet{2013ApJ...767...50M}, as the distinction is not particularly sharp. Fig.~\ref{fig:fpass} shows the passive fraction, along with the SDSS/GALEX data and another model from the literature. Neither of the models nor the data agree with one another. An increase in the passive fraction at low mass would ease the discrepancy with the passive SMF, but make the discrepancy with the SFMS SMF worse. It is not clear that a globally acceptable fit could be found for both the SFMS and passive SMFs when taking the data at face value.

\begin{figure} 
\includegraphics[width=0.45\textwidth]{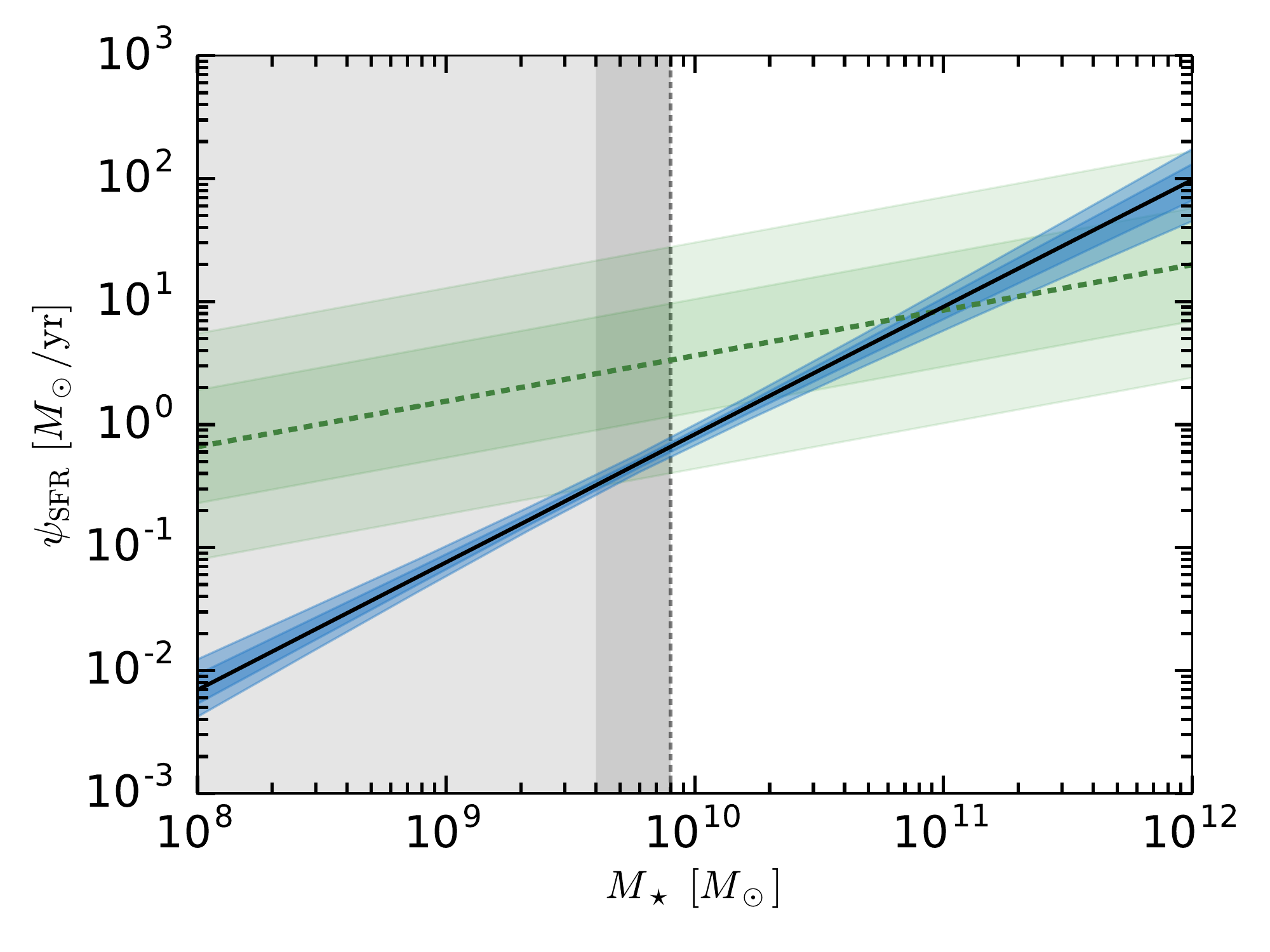}\vspace{-1em}
\caption{The mean star-formation rate as a function of stellar mass. The blue bands show 68\% and 95\% credible intervals for our model from the MCMC. The green dashed line shows the the best-fit \citet{2013MNRAS.431..648W} relation in the redshift range $z=[0.2, 0.5]$ (corresponding to $\alpha = -3.14$, $\beta = 0.37$) along with the reported errors on the fit (green bands). The grey shaded regions show stellar masses that lie below the completeness limit of some (or all) of the data used by \citet{2013MNRAS.431..648W} to fit their model.} \vspace{-1em}
\label{fig:sfrmstarbounds}
\end{figure}

Residual systematic effects could also be present in the data. Fig.~4 of \citet{2013ApJ...767...50M} shows the total SMF from SDSS/GALEX plotted alongside four previous determinations of the SMF from the literature. The difference between the curves is several times the errorbars in most cases, suggesting that at least some of the five datasets are inconsistent with one another.

Finally, while the SDSS/GALEX errorbars shown in Fig.~\ref{fig:smf} are small, there still appears to be relatively little scatter in the positions of neighbouring datapoints, indicating that the errors are correlated. We were unable to find information in \cite{2013ApJ...767...50M} to quantify this (e.g. a correlation matrix), but if the correlation is strong, this will substantially reduce the amount of independent information in the data, and thus the significance of the discrepancy with our model prediction.

Fig.~\ref{fig:sfrmstarbounds} shows the best-fit model for the mean SFR relation of the star-forming main sequence (Eq.~\ref{eq:meansfms}), along with 68\% and 95\% credible intervals from our MCMC chains, and for a previous fit of a power-law relation to other data by \citet{2013MNRAS.431..648W} (in the redshift bin $z = [0.2, 0.5]$). Roughly following the trend with redshift found in \citet{2013MNRAS.431..648W}, one would expect an extrapolation of their model to $z=0$ to reduce its amplitude, but leave the slope relatively unchanged. The slope of our model is clearly much steeper. The comparison should only be performed above the completeness limit of the data that they used however, as denoted by the grey bands in Fig.~\ref{fig:sfrmstarbounds}. The errors on their fit are relatively large, and essentially cover the posterior of our relation above the completeness limit. \citet{2013MNRAS.431..648W} do report that their best-fit model has a reduced $\chi^2$ of 0.2 though, which likely indicates that the errorbars are overestimated.

\begin{figure} 
\hspace{-2em}\includegraphics[width=0.52\textwidth]{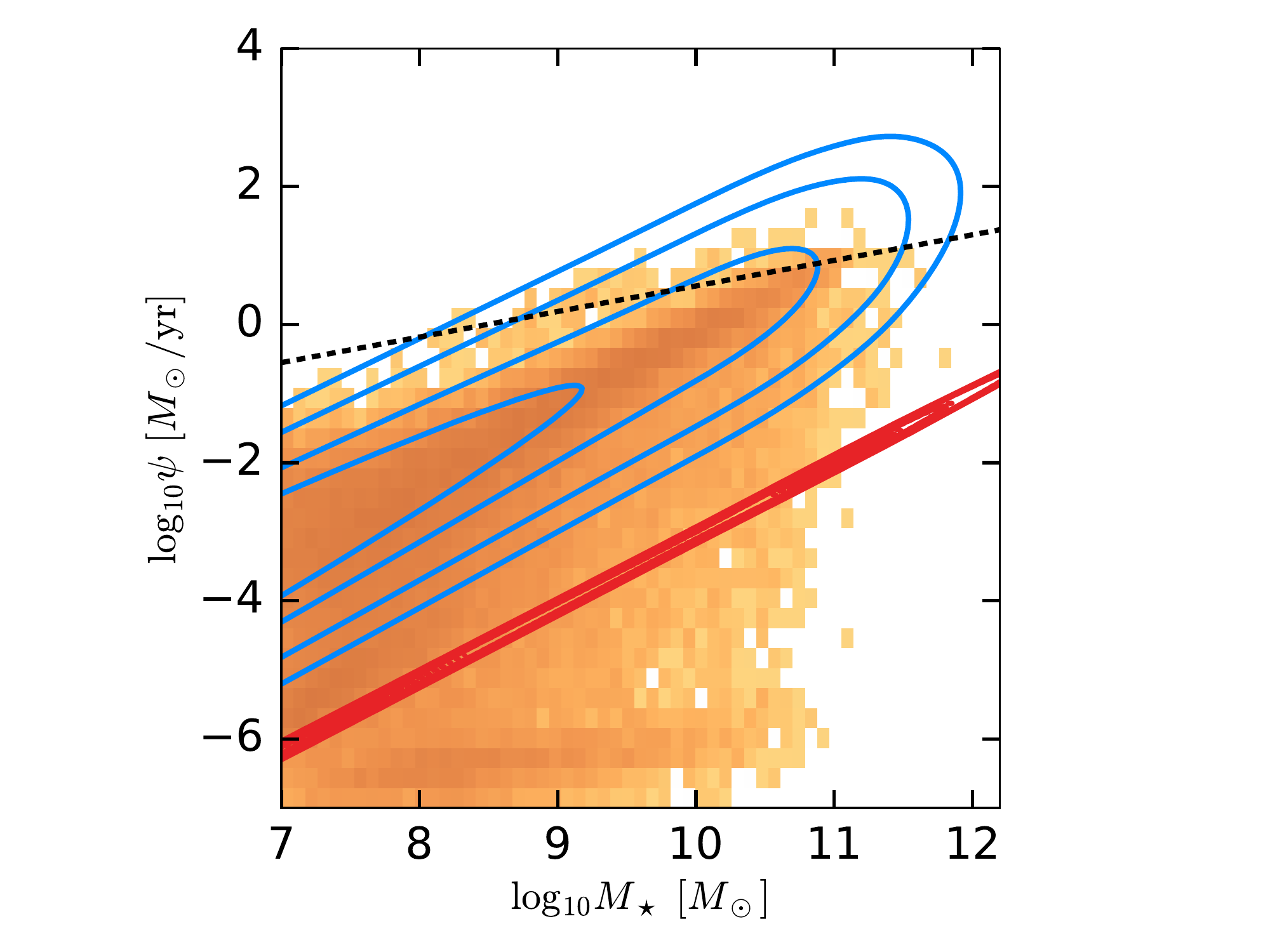}
\vspace{-1em}
\caption{Predicted number density distribution in the $\mstar-\sfr$ plane, $dn/d\log \mstar d\log\sfr$, for star-forming main sequence (blue) and passive (red) galaxies. The model shown is the best-fit model to the GAMA and NVSS optical/radio luminosity function data. Contours are shown at values of $\{10^{-3}, 10^{-4}, 10^{-6}, 10^{-8}\}$ Mpc$^{-3}$ (SFMS) and $\{10^{-4}, 10^{-6}, 10^{-8}\}$ Mpc$^{-3}$ (passive) respectively. The density plot (orange) shows the same quantity measured from the \citet{Guo:2010ap} semi-analytic mock catalogue at $z=0$. This simulation was not used in the fits, except to define the optical magnitude parameters in Eq.~\ref{eq:mag_fit}. The dashed line is the same as in Fig.~\ref{fig:sfrmstarbounds}.}
\label{fig:sfrmstar}
\end{figure}

\begin{figure*} 
\hspace{-2.5em}
\includegraphics[width=1.03\textwidth]{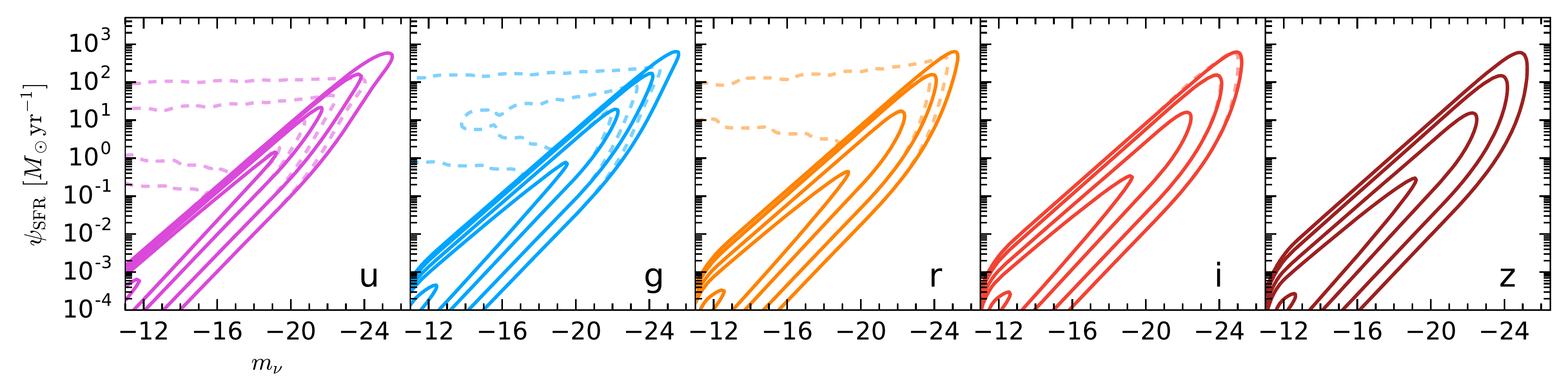}
\caption{Joint luminosity function of galaxies on the SFMS at $z=0$, for a tracer of the SFR (e.g. radio continuum at 1.4 GHz) and five optical bands. The solid contours show the luminosity function without dust attenuation in the optical, while the dashed lines show the result with attenuation. The values of the contours are $dn/d\log\sfr dm_\nu = \{ 10^{-8}, 10^{-6}, 10^{-4}, 10^{-3}, 10^{-2} \}$ Mpc$^{-3}$ (from outermost to innermost).}
\label{fig:jointlf}
\end{figure*}

The number density distribution in the $\sfr - \mstar$ plane is plotted in Fig.~\ref{fig:sfrmstar} for both the SFMS and passive population. These are compared with the distribution at $z=0$ from the \citet{Guo:2010ap} simulation. The SFMS aligns very well with the one found in the simulation, with essentially the same slope and a similar width. Our model appears to overestimate the density of galaxies at high $\sfr$ and $\mstar$, but this could also be due to under-sampling in the simulation (which has a relatively small box size of 62.5 $h^{-1}$Mpc). The passive distribution is entirely different from the one in the simulation however; the simulated one has no sequence-like structure, while our best-fit model prefers a very narrow sequence. One can imagine that a much larger value of $\sigma_{\rm pass}$ would have been able to approximately reproduce the simulated result if this was preferred by the data, but it is not. As discussed in Sect.~\ref{sec:analysis}, part of the reason for this preference might be the need to reproduce a small, narrow feature in the optical LFs. The various semi-analytic models for the passive population also greatly differ from one another (see Sect.~\ref{sec:sfrmstar}), and so the one that we chose (and/or the one in the simulation) may simply be inconsistent with the data. This could also cause the large discrepancy with the passive SMF that was seen in Fig.~\ref{fig:smf}.

In summary, our model appears to reproduce previous results reasonably well, yielding qualitatively good fits to the optical and radio LFs, reasonable overlap with a previous determination of the star-forming main sequence, and only a mild underestimate of the stellar mass function of star-forming galaxies. The passive sequence of the best-fit model appears to be strongly discrepant with simulations and some previous results however (although we have noted that different simulations give a broad spread of predictions for the shape of the passive population in the SFR-$\mstar$ plane). A detailed investigation to construct a more accurate model for the passive population is left for future work.

\subsection{Joint luminosity functions}

We now give an example of the kind of multiple-tracer prediction that the model is designed to calculate. Given two samples of galaxies in different wavebands, it is useful to quantify what fraction of the galaxies can be detected in both bands simultaneously. This can be used to predict how efficiently sources can be cross-matched between surveys, for example, or for calculating correlated shot noise in the cross-correlation of the surveys. The basic quantity required for these calculations is the joint luminosity function, ${dn(L_1, L_2)/d\log L_1 d\log L_2}$. As an example, we will calculate the joint luminosity function for one optical band and the 1.4 GHz radio continuum emission from star-forming galaxies. This can be written as
\bea
\frac{dn(L_{\rm rad}, m^{\rm obs}_\nu)}{d L_{\rm rad} dm^{\rm obs}_\nu} &=& \int d\sfr \int d\mstar \int d \mhalo \frac{dn(\mhalo)}{d\mhalo} \nonumber\\ 
&& \times \,\, p(L_{\rm rad}, m^{\rm obs}_\nu | \mstar, \sfr) \nonumber\\
&& \times \,\, p(\sfr | \mstar)\cdot p(\mstar | \mhalo). \label{eq:jointlum}
\eea
All of these distributions were defined above, with the exception of the joint luminosity/magnitude pdf, which we model as
\be
p(L_{\rm rad}, m^{\rm obs}_\nu | \mstar, \sfr) = p(L_{\rm rad} | \sfr) \cdot p(m_\nu^{\rm obs} | \mstar, \sfr).
\ee
This assumes that the two types of emission are linked only by their common dependence on $\sfr$ and $\mstar$, i.e. that they are independent when we condition on SFR and stellar mass. We have also used the fact that the radio luminosity depends only on SFR. In fact, our model assumes a deterministic relationship between SFR and $L_{\rm rad}$ (Eq.~\ref{eq:sfradio}), so that
\be
p(L_{\rm rad} |\, \sfr) = \delta^{\rm D}(\sfr - \sfr^{\rm rad}(L_{\rm rad})),
\ee
where $\delta^{\rm D}$ is the Dirac delta function. Eq.~\ref{eq:jointlum} then simplifies to
\bea
\frac{dn(L_{\rm rad}, m^{\rm obs}_\nu)}{d \log L_{\rm rad} dm^{\rm obs}_\nu} = L_{\rm rad} \int d\mstar \frac{dn(\mstar)}{d\mstar}\, p(\sfr^{\rm rad}(L_{\rm rad}) | \mstar) \nonumber\\
\times \,\, p(m_\nu^{\rm obs} | \mstar, \sfr^{\rm rad}(L_{\rm rad})),
\eea
where we have also changed variables to $\log L_{\rm rad}$.

With the joint luminosity function in hand, it is straightforward to evaluate the number density of galaxies that are detected by both optical and radio surveys. Assuming luminosity and magnitude limits $L_*$ and $m_*$ respectively, we have
\begin{align}
n(L_{\rm rad} \ge L_*, m^{\rm obs}_\nu \le m_*) 
= \int_{\log L_*}^\infty \int^{m_*}_{-\infty} \frac{dn(L, m_\nu)}{d \log L\, dm_\nu}\, d\log L\,\, dm_\nu.
\end{align}
The fraction of optically-detected galaxies that also have radio counterparts (and vice versa) is then
\bea
f({{\rm rad}\, \in\, {\rm opt}}) &=& \frac{n(L_{\rm rad} \ge L_*, m^{\rm obs}_\nu \le m_*)}{n(L_{\rm rad} \ge 0, m^{\rm obs}_\nu \le m_*)} \\
f({{\rm opt}\, \in\, {\rm rad}}) &=& \frac{n(L_{\rm rad} \ge L_*, m^{\rm obs}_\nu \le m_*)}{n(L_{\rm rad} \ge L_*, m^{\rm obs}_\nu \le \infty)},
\eea
where the denominators are simply the total number density of optical/radio galaxies in each survey respectively, regardless of whether they have radio/optical counterparts or not.

Fig.~\ref{fig:jointlf} shows the joint luminosity function for optical bands and a general direct SFR tracer (like radio continuum) as predicted by our model, for the best-fit parameter values described above. As expected, galaxies that are most actively star-forming are brightest in the optical, although there is not a 1:1 correspondence even in the case where dust attenuation is neglected. This is simply a consequence of the optical magnitudes also tracing the stellar mass of the galaxies; the spread in $\mstar$ for fixed $\sfr$ and the scatter in the optical magnitude pdf broaden the joint luminosity function. The correlation is tighter in the bluer bands, where the emission is dominated by younger stars that trace recent star formation.

With attenuation included, the picture becomes more complicated. For the bluest bands, in which attenuation is the strongest, a branch of high-SFR galaxies appears that spreads broadly down to fainter optical magnitudes. The spreading effect is reduced at lower SFR however, mostly because our model of attenuation scales with stellar mass; since $\sfr$ also scales with $\mstar$, lower SFR corresponds, on average, to lower stellar mass and thus weaker attenuation. Note that the number density of galaxies in this branch is relatively low, and that the main effect of attenuation is the shifting of the bright end of the joint luminosity function to fainter optical magnitudes. 


\section{Discussion} \label{sec:discussion}

We have presented a modular, analytic statistical model that connects host dark matter halo properties to bulk properties of the galaxies that reside within them, which are then linked to observables such as galaxy luminosity across a range of bands. This model was motivated by the existence of various scaling relations between these quantities, and the need to model multiple tracer populations across wavelength regimes when analysing forthcoming multi-survey large-scale structure datasets.

We showed that conditional distributions representing various scaling relations -- many of which were already known in the literature -- can be combined in a consistent way to simultaneously model the luminosity functions of low redshift galaxies in the optical and radio, with good precision. Consistency is achieved by jointly fitting the model parameters to multi-wavelength data, instead of combining relations that were separately fitted to various datasets. We also reported on the uncertainties and correlations between model parameters that resulted from this fitting procedure, using the results of an MCMC sampling procedure.

As presented here, our model predicts only a limited set of galaxy properties that are important for future surveys. We did not attempt to model (e.g.) the neutral gas content or emission from nuclear activity, even though these will be important for radio surveys for example. We posit that many such properties can be included as straightforward extensions to the model, by adding suitable conditional distributions to the probabilistic graph shown in Fig.~\ref{fig:model}. This will generally involve finding appropriate scaling relations between galaxy properties and then promoting them to (parametrised) statistical distributions.

Many such scaling relations have been seen observationally or in simulations, or result from simple physical arguments. Properties of active galactic nuclei (AGN) such as their radio and X-ray luminosity may be modelled through scaling relations involving the mass of their central black hole, stellar mass, or halo mass, for example \citep{2002ApJ...578...90F, 2011MNRAS.413..957F, 2012ApJ...746...90A}. 
Relatively simple models of the neutral hydrogen (HI) content of galaxies also exist \citep[e.g.][]{2009ApJ...698.1467O, 2010MNRAS.407..567B, 2014JCAP...09..050V, 2016arXiv160701021P}, often involving a scaling between HI mass and halo mass only. Aside from using the simple star-forming/passive categorisation to assign disk/elliptical labels to the galaxies, more sophisticated morphological properties can also be modelled using scaling relations \citep[e.g. see the treatment in][]{2016arXiv160605354S}. Adding these properties into the model, plus luminosity relations for other bands, is left for future work, but it should be clear that extensions like this are supported in a natural way.

While our model does explicitly include per-band optical magnitudes, including dust attenuation, there is some further structure in the relationships between different bands that is not modelled -- namely, colours. Galaxy colours are commonly used as selection criteria in surveys, as structures exist in colour-magnitude diagrams that can reliably distinguish between different types of galaxy. The existence of a red sequence in the colour-magnitude diagram \citep{2004ApJ...608..752B} can be used to identify the large red galaxies that typically lie at the centre of galaxy clusters, for example \citep[e.g.][]{2014ApJ...785..104R}. Our model fails to reproduce the red sequence however. This is because the optical pdfs (Eq.~\ref{eq:opticalpdf}) are assumed independent; the magnitudes in different bands are linked only through the mutual dependence of the mean relations (Eq.~\ref{eq:mag_fit}) on stellar mass and SFR. Sequences in the colour-magnitude diagram {\corr also require correlations in the {\it scatter} between two bands, however. This could potentially be implemented by replacing Eq.~\ref{eq:opticalpdf} with a multivariate lognormal pdf for all bands, with $\sigma_m^{(\nu)}$ replaced by a full covariance matrix, thus at least allowing (log-)Gaussian correlations in the scatter.}

While sufficient for some applications, this would also limit the model. Real galaxies have complex optical SEDs that depend on a host of other factors, in addition to bulk properties like stellar mass and SFR. The star-formation history of the galaxy and its initial mass function (IMF) are two such ingredients, both of which were used in the semi-analytic simulations from which our optical magnitude relations were calibrated. In future work, it would be desirable to remove the dependence of this part of the model on SAMs, and all of the assumptions that underlie them. One way of achieving this would be to build the optical magnitudes from bandpass integrals over distributions of template galaxy spectra, such as those derived from the stellar population synthesis models of \citet{2003MNRAS.344.1000B}. This method has been used in a number of galaxy mock catalogue generation algorithms, {\corr as well as in MCMC analyses of galaxy spectra \citep[e.g.][]{2017ApJ...837..170L}.} The downside is that it introduces significant extra complexity, still requires a number of assumptions (e.g, about SFR history), and destroys the analytic nature of our model. A middle ground might be to use sets of empirical galaxy spectrum templates, such as those generated by some photometric redshift estimation methods \citep[e.g.][]{2015ApJ...813...53M}, although this also adds significant complexity. 

The model is incomplete in another important respect -- we have calibrated it using only $z \simeq 0$ observations. The redshift dependence of several of the constituent conditional distributions has been studied previously (e.g. see \citet{2010ApJ...717..379B} for the redshift dependence of the stellar mass-halo mass relation), and so redshift-dependent parametrisations are readily available. These simply add more parameters to the model that can be constrained with additional data at other redshifts. Other components are new to this work, however. We have tentatively confirmed that some of these components have redshift dependences that are amenable to simple parametrisations, by performing fits on semi-analytic simulations (e.g. for Eq.~\ref{eq:mag_fit}). A full treatment, including constraining the model from higher-redshift data, is left to future work.\footnote{Ideally, this would result in every component having a redshift dependence that is well-described by simple, few-parameter extensions to the model. In the worst case is should still be possible to find a new set of best-fit parameters for every redshift slice, however.}

{\corr In future, the model will also need to account for clustering observables. This will likely require several modifications to the early parts of the chain of conditional distributions that link halo mass to galaxy type, stellar mass, and star-formation rate. The current setup would not be able to explain the observed SFR-dependence of clustering, for instance \citep[c.f.][]{2013ApJ...778...93T, 2016MNRAS.457.4360Z, 2017ApJ...838...87C}. Models that are able to successfully reproduce observed clustering signals generally divide galaxies into `centrals' and `satellites', and permit more than one galaxy per host halo. These features could be implemented by adding a conditional distribution for the satellite mass function after the halo mass function (see Fig.~\ref{fig:model}), and allowing later steps of the chain to also be conditioned on the central vs. satellite classification. Scaling relations that differentiate between centrals and satellites in (e.g.) the stellar mass-halo mass relation already exist in the literature \citep[e.g.][]{2010ApJ...710..903M}, and would be straightforward to adopt, at the expense of adding several additional parameters. Later parts of the chain that depend only on stellar mass and SFR would likely remain unchanged. A proof of concept is left to future work, however.}

We have also neglected several observational effects that are important in modelling the observed distribution of galaxy luminosities. A small fraction of galaxies are strongly lensed, for example, which enhances their apparent luminosity. This causes a magnification bias \citep{2006glsw.conf.....M}, which can be especially important at the bright end of the luminosity function, where intrinsically bright objects may be rarer than the lensed population of fainter objects. As such, our luminosity functions should be convolved with a lensing magnification pdf \citep[e.g.][]{2013PhRvD..88f3004M} to account for this effect.

{\corr Finally, we note that a shortcoming of this kind of modelling is the inevitable proliferation of parameters. As the model is extended to account for increasingly complex observables, new model components (and therefore new parameters) will need to be added. The question is whether the growing parameter space will eventually become unwieldy, at least compared with just using SAMs or other more physics-based modelling approaches. This very much depends on what the model will be used for, and how the parameters will be constrained. In this paper, we fixed a number of the optical magnitude parameters, based on the outputs of a SAM. This `calibration' approach helps to keep the number of degrees of freedom in check while allowing the model complexity to be increased significantly, and is legitimate if one has confidence that the input parameter values (whether taken from a theoretical model or external data) are sufficiently realistic and well constrained to be fixed in this way. This is effectively done by SAMs and hydrodynamical simulations too, which often have hidden `fixed' degrees of freedom, such as libraries of stellar templates, fixed stellar population models, and sub-grid physics.}

{\corr In general, we expect that current galaxy surveys and simulations are sufficiently information-rich to allow significantly more model parameters to be constrained, or sensibly fixed a priori, than were considered here. Some parameters will likely remain ill-constrained, or degenerate with one another, but this is not necessarily problematic if the target observables can still be calculated sensibly when these are marginalised over. The question is then whether the user of the model finds the added complexity acceptable and/or useful, or whether other approaches are more appropriate for their intended application. It is hard to say exactly how much more complexity would be required to achieve our goal of accurately modelling the 1-point and 2-point functions of several different tracers, but we are hopeful that it can be managed by adding only a few additional components, as discussed above.}

{\corr This process will be aided by the modular nature of the model, which allows us to easily compare several alternatives for each of the constituent conditional distributions.} Through statistical tests, such as Bayesian model selection procedures, one can then choose components that most faithfully describe various observations, while penalising overly-flexible models that lack predictive power. These can then be swapped into the default configuration of the model, without requiring it to be completely rebuilt.

The type of model we have presented here fills a gap between detailed, computationally-expensive galaxy modelling methods like semi-analytics and hydrodynamical simulations on the one hand, and single-population halo models (e.g. HOD models) on the other. It has sufficient structure to allow consistent modelling of multiple galaxy populations across multiple wavelengths, but is simple enough that it can be explored analytically, allowing MCMC analyses to be performed without extensive computational resources. An open-source reference implementation, written in {\tt Python}, and making use of the {\tt numpy} and {\tt scipy} \citep{scipy} packages, is available from \url{https://github.com/philbull/ghost}.

\vspace{1em}
{\sl Acknowledgements:} I am grateful to O.~Dor\'{e}, B.~Hensley, E.~Huff, A.~Merson, P.~Serra, M.~Viero, H.-Y.~Wu, {\corr and an anonymous referee} for valuable discussions, comments, and encouragement, and to A.~Brown for valuable work on a summer project involving this model. I gratefully acknowledge use of computing resources at the University of Oslo. PB's research was supported by an appointment to the NASA Postdoctoral Program at the Jet Propulsion Laboratory, California Institute of Technology, administered by Universities Space Research Association under contract with NASA.

\FloatBarrier

\setcounter{figure}{0}
\renewcommand{\thefigure}{A}
\begin{figure*} 
\includegraphics[width=0.74\textheight]{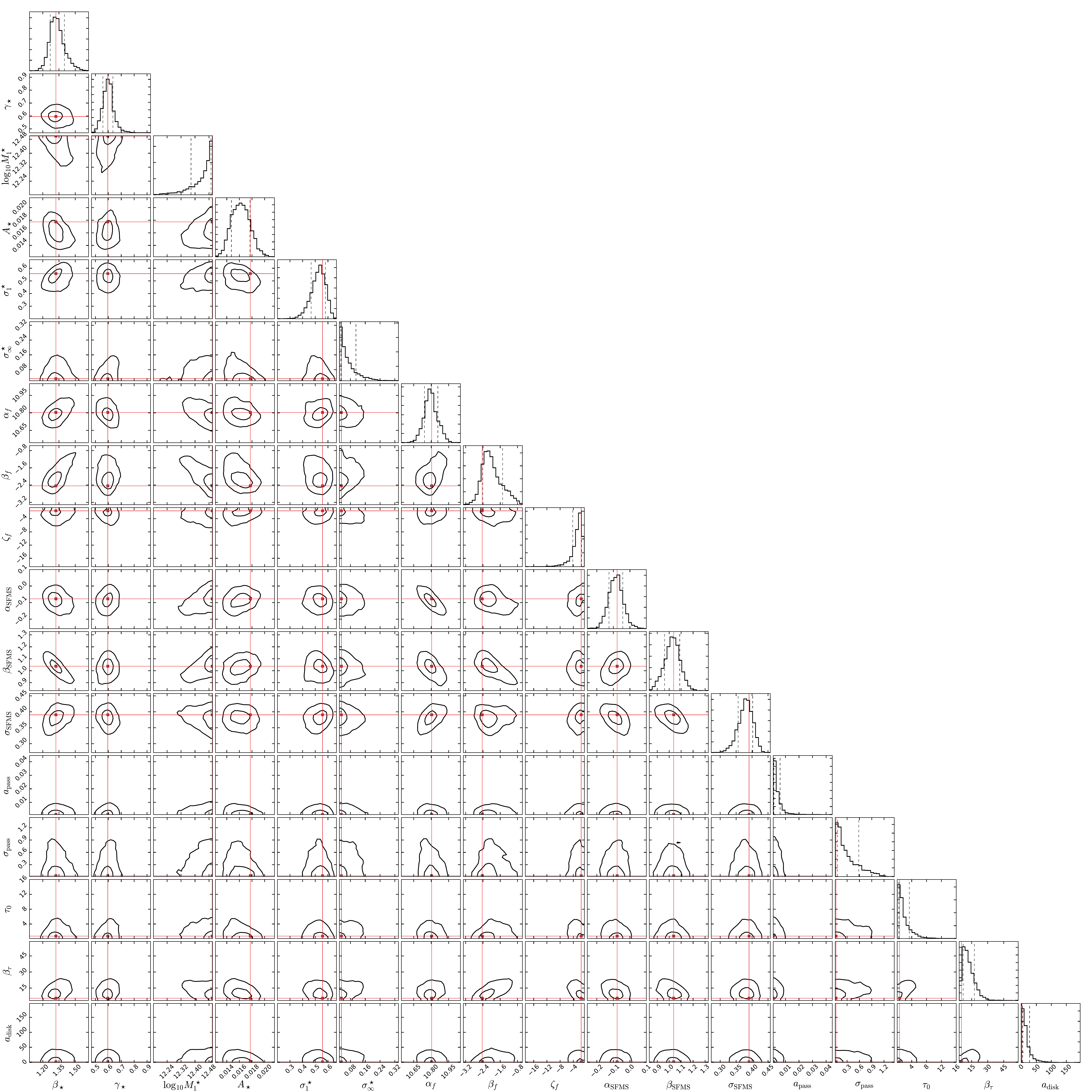}
\caption{2D posterior distributions of the 17 parameters that were varied in the MCMC. The contours are $1$ and $2\sigma$ bounds, and the marker shows the best-fit model. The 1D marginal distribution for each parameter is shown on the diagonal. Some of the parameters with strongly non-Gaussian 1D marginal distributions are restricted by their prior bounds (e.g. $\sigma_\infty^\star$), while others simply prefer a logarithmic distribution (e.g. $a_{\rm disk}$). We used the {\tt corner} software \citep{corner} to make this plot.}
\label{fig:triangle}
\end{figure*}

\balance

\bibliographystyle{aa}
\bibliography{ghmodel}

\end{document}